\documentstyle[eqsecnum,prd,aps,epsf]{revtex}
\draft

\begin{document}
\newcommand{\beq}{\begin{equation}}
\newcommand{\eeq}{\end{equation}}
\newcommand{\beqn}{\begin{eqnarray}}
\newcommand{\eeqn}{\end{eqnarray}}
\newcommand{\beaa}{\begin{eqnarray*}}
\newcommand{\eeaa}{\end{eqnarray*}}
\newcommand{\dis}{\displaystyle}
\newcommand{\Lie}{\mbox{\pounds}}
\newcommand{\pa}{\partial}
\newcommand{\na}{\nabla}
\newcommand{\vp}{\varphi}
\def\Madm{M_{\rm ADM}}
\def\Jadm{J_{\rm ADM}}
\def\Mchi{M_{\rm \chi}}
\def\Mmix{M_{\rm mix}}
\def\MK{M_{\rm K}}
\newcommand{\rhoH}{\rho_H}
\newcommand{\HG}{{\cal H}_G}
\newcommand{\CG}{{\cal C}_G}

\def\2pi{2\pi}
\def\zero{\hbox{$_{(0)}$}}
\def\bL{\hbox{$\,{\cal L}\!\!\!\!\!-$}}
\def\bI{\hbox{$\,I\!\!\!\!-$}}
\def\zeroD{\stackrel{{(0)}}{D}}
\def\fourR{\stackrel{{(4)}}{R}}
\def\zeroDelta{\stackrel{(0)}{\Delta}}
\def\two{\hbox{$^{(2)}$}}
\def\three{\hbox{$^{(3)}$}}
\def\four{\hbox{$^{(4)}$}}
\def\tn{\hbox{$^{(n)}$}}
\newcommand{\gab}{g_{\alpha\beta}}
\newcommand{\gabu}{g^{\alpha\beta}}
\newcommand{\gabd}{g_{\alpha\beta}}
\newcommand{\gmabu}{\gamma^{ab}}
\newcommand{\gmabd}{\gamma_{ab}}
\newcommand{\tgmabu}{\tilde\gamma^{ab}}
\newcommand{\tgmabd}{\tilde\gamma_{ab}}
\newcommand{\tgamma}{\tilde\gamma}
\newcommand{\thh}{\tilde h}
\newcommand{\tf}{\tilde f}
\newcommand{\piabu}{\pi^{ab}}
\newcommand{\piabd}{\pi_{ab}}
\newcommand{\piab}{\pi_a{}^b}
\newcommand{\tpiabu}{\tilde{\pi}^{ab}}
\newcommand{\tpiabd}{\tilde{\pi}_{ab}}
\newcommand{\tpi}{\tilde{\pi}}
\newcommand{\qab}{q_\alpha{}\!^\beta}
\newcommand{\qbc}{q_\beta{}\!^\gamma}
\newcommand{\qabu}{q^{\alpha\beta}}
\newcommand{\qabd}{q_{\alpha\beta}}
\newcommand{\Tabd}{T_{\alpha\beta}}
\newcommand{\Tabu}{T^{\alpha\beta}}
\newcommand{\Tab}{T^\alpha{}\!_\beta}
\newcommand{\Gab}{G^\alpha{}\!_\beta}
\newcommand{\Gabd}{G_{\alpha\beta}}
\newcommand{\Gabu}{G^{\alpha\beta}}
\newcommand{\Rab}{R^\alpha{}\!_\beta}
\newcommand{\dSab}{dS_{\alpha\beta}}
\newcommand{\dSa}{dS_{\alpha}}
\newcommand{\QK}{Q_{\rm K}}
\newcommand{\dl}{\delta}
\newcommand{\dlab}{\dl^\alpha{}\!_\beta}
\newcommand{\dlabu}{\dl_\alpha{}\!^\beta}
\newcommand{\dlgab}{\delta g_{\alpha\beta}}
\newcommand{\Dlgab}{\Delta g_{\alpha\beta}}
\newcommand{\Dlgabu}{\Delta g^{\alpha\beta}}
\newcommand{\Dl}{\Delta}
\newcommand{\dw}{\ \,}
\newcommand{\dww}{\ \,\,}
\newcommand{\Lag}{{\cal L}}

\title{ 
Deriving formulations for numerical computation of
binary neutron stars in quasicircular orbits
}
\author{
Masaru Shibata$^1$, K\=oji Ury\=u$^2$, and John L. Friedman$^3$ }

\address{ 
$^1$Graduate School of Arts and Sciences,~University of 
Tokyo,\\
Komaba, Meguro, Tokyo 153-8902, Japan \\
 $^2$ SISSA, Via Beirut 2/4, 34014 Trieste, Italy\\
$^3$ Department of Physics,
University of Wisconsin-Milwaukee, P.O. Box 413, 
Milwaukee, WI 53201\\
}
\maketitle

\begin{abstract}
Two relations, the virial relation $M_{\rm ADM}=M_{\rm K}$
and the first law in the form $\delta M_{\rm ADM}=\Omega \delta J$, 
should be satisfied by a solution and 
a sequence of solutions describing binary compact objects 
in quasiequilibrium circular orbits. Here, $M_{\rm ADM}$, 
$M_{\rm K}$, $J$, and $\Omega$ are the ADM mass, Komar mass, 
angular momentum, and orbital angular velocity, respectively.  
$\delta$ denotes an Eulerian variation. 
These two conditions restrict the allowed formulations 
that we may adopt. First, we derive relations between 
$M_{\rm ADM}$ and $M_{\rm K}$ and between 
$\delta M_{\rm ADM}$ and $\Omega \delta J$ for 
general asymptotically flat spacetimes. 
Then, to obtain solutions that satisfy the virial relation and 
sequences of solutions that satisfy the first law at least approximately, 
we propose a formulation for computation of 
quasiequilibrium binary neutron stars in general relativity. 
In contrast to previous approaches in which a part of
the Einstein equation is solved, in the new formulation, 
the full Einstein equation is solved 
with maximal slicing and in a transverse gauge for 
the conformal three-metric. Helical symmetry is imposed 
in the near zone, while in the distant zone, a waveless condition
is assumed. We expect the solutions obtained in this formulation 
to be excellent quasiequilibria as well as
initial data for numerical simulations of 
binary neutron star mergers. 
\end{abstract}
\vskip 2mm
\maketitle

\section{Introduction}
\label{sec:intro}

A detailed theoretical understanding of evolution of close
binary neutron stars is one of the most important goals 
in general relativity, since they are 
promising sources of gravitational waves for laser interferometric 
gravitational wave detectors such as LIGO, TAMA, GEO, and VIRGO
\cite{LIGO,TAMA}. For the inspiral phase in which the 
orbital separation $a$ is much larger 
than the radius $R$ of a neutron star, the orbital velocity is
much smaller than the speed of light, and finite-size effects of
neutron stars may be ignored. Thus, 
a post Newtonian study together with the point particle approximation 
is appropriate \cite{blanchet}.
For $a/R \alt 4$, however, the post Newtonian and point
particle approximations break down and numerical study is required
to take into account the effect of tidal deformation of each star and
full effects of general relativity. 
The procedure to be adopted for such close orbits up to the merger 
is (i) to compute a quasiequilibrium circular orbit 
at a distant orbit with $a \alt 4R$ and $a \agt 10M$ for which
ratio of a radial approaching velocity to the orbital one
will be small (less than 1\%) \cite{SU01},
and then (ii) to perform a numerical relativity simulation adopting
a distant quasiequilibrium with $a \agt 10M$
as the initial condition \cite{SU,MGS}. 
In this paper, we focus on the formulation for computation of the 
quasiequilibrium in circular orbits. 

So far, the quasiequilibrium states of binary neutron stars have
been widely computed in the so-called conformal flatness approximation 
(or Isenberg-Wilson-Mathews formalism) \cite{IWM,bcsst97,BGM,MMW,UE},
in which the conformal three-metric is assumed to be flat. 
The solution in this formulation satisfies the 
constraint equations of general relativity and, hence, it
is fully general relativistic for the initial value problem.
However, it is only an approximate {\it quasiequilibrium} solution for
compact binaries, since conformally nonflat
parts of the three-metric are not vanishing for the quasiequilibrium
binaries. Thus, such approximation produces a systematic error 
of magnitude $\sim (M/a)^2$ 
for the solution of quasiequilibrium configurations \cite{SU01}. 
The systematic error is also included in 
gravitational waves computed from the conformal flat data
of the quasiequilibrium binary \cite{SU01,DBSSU}
and in the numerical results of fully general relativistic simulations
started from initial conditions 
of the conformally flat quasiequilibria \cite{SU}. 

Formulations for computation of binary compact objects in
quasiequilibrium circular orbits 
with a conformally nonflat three-metric have been proposed
by several authors (e.g., \cite{BCT,Laguna,BGGN04,SG04,bd92d94,Price,SU01}). 
A promising approach to this problem is to assume a helical 
Killing symmetry for the spacetime \cite{bd92d94,Price}. 
In this case, however, the solution contains standing gravitational
waves in the whole spacetime and the 
averaged energy density of gravitational waves falls off as
$r^{-2}$ where $r$ denotes a radial coordinate, resulting in an 
asymptotically nonflat spacetime. Thus, the solution 
obtained in such formulation is not physical in the distant
wave zone, although the solution in the near zone and
in a local wave zone would describe a realistic spacetime 
of binary compact objects.

In this paper, we consider general relativistic formulations 
for computation of the quasiequilibrium circular orbits assuming 
that the spacetime is asymptotically flat. 
First, we require that the following two conditions should be
satisfied for a solution and a sequence of the solutions
of quasiequilibrium states: 
\begin{itemize}
\item A quasiequilibrium solution that is stationary in the 
corotating frame should satisfy a virial relation associated 
with the equality  
\beq
M_{\rm ADM}=M_{\rm K},
\eeq
of the ADM and Komar masses defined in Sec. III. 
\item Binary compact objects inspiral adiabatically as a result of
gravitational wave emission, conserving baryon mass, entropy,
and vorticity. Thus, along a sequence of 
quasiequilibrium solutions, the first law should be satisfied.
Here, the first law is written in the form
\beq
\delta M_{\rm ADM} = \Omega \delta J, \label{first}
\eeq
where $\delta M_{\rm ADM}$ and $\delta J$ are infinitesimal differences of
the ADM mass and angular momentum along a quasiequilibrium sequence, and 
$\Omega$ is an orbital angular velocity.
\end{itemize}
These two conditions are likely to be satisfied for
binary neutron stars in nature. Thus, we should adopt 
a formulation which provides a solution and a sequence 
of the solutions that satisfy two conditions at least approximately. 

Based on this motivation, in this paper, we first derive
relations for the differences, $M_{\rm ADM}-M_{\rm K}$ and
$\delta M_{\rm ADM} - \Omega \delta J$, in arbitrary asymptotic
flat spacetimes.
The condition that the differences vanish  
can be used to restrict formulations that we can adopt. 
Using these conditions, several possible candidates for the 
formulations emerge. 
Among them, we propose a formulation in which helical symmetry 
is imposed only in the near zone 
instead of in the whole spacetime. 
Specifically, we impose a mixed condition; a helical symmetry
condition in the near zone and a waveless condition 
in the distant zone. To fix the gauge, we adopt the 
maximal slicing condition and a transverse gauge condition for 
the upper component of the conformal three-metric. In this case, all 
components of the Einstein equation reduce to 
elliptic equations as in the post Newtonian approximation. 
This implies that no standing waves appear in the wave zone, 
although in the near zone, gravitational-wave-like components are 
present. We expect the solutions obtained in this formulation 
to be excellent quasiequilibria as well as
initial data for numerical simulations of binary 
neutron star mergers. 

The paper is organized as follows. 
In Sec. II, we describe the basic equations for quasiequilibria. 
In Sec. III, we derive a relation between 
$M_{\rm ADM}$ and $M_{\rm K}$ for arbitrary formulation and 
clarify the condition for the formulation that its solution satisfies 
the virial relation $M_{\rm ADM}=M_{\rm K}$. 
In Sec. IV, we derive a relation for the difference, 
$\delta M_{\rm ADM} - \Omega \delta J$, and 
clarify the conditions on the formulation for which a sequence of
solutions satisfies the first law. 
In Sec. V, we propose formulations whose solutions and
sequences of solutions approximately satisfy 
the virial relation and the first law. 
Section VI is devoted to a summary. 

Throughout this paper, we use geometrical units with $G=1=c$. 
Spacetime indices are Greek, spatial indices 
Latin, and the metric signature is $-+++$. Readers familiar with abstract 
indices can regard indices early in the alphabet as abstract, while 
$i, j, k, \cdots$ are concrete, associated with a chart $\{x^i\}$. 
If $S$ is a 2-surface in a 3-space $\Sigma$ and $\epsilon_{abc}$ is the 
volume form on $\Sigma$ associated with a 3-metric $\gamma_{ab}$, we write 
$dS_a = \epsilon_{abc}dS^{bc}$; for $S$ a surface of constant $r$,
$dS_a = \na_a r \sqrt{\gamma} d^2 x$.

\section{Formulation}

\subsection{3+1 formalism}
\label{sec:3+1form}

Let $\Sigma_t$ be a family of spacelike hypersurfaces, labeled by a 
time function $t$.  Let $t^\alpha$ be a vector field transverse to $\Sigma_t$
for which $t^\alpha\na_\alpha t=1$, and denote by $\alpha$ and $\beta^\alpha$ 
a nonvanishing lapse and a shift vector, respectively, with    
\beq
 t^\alpha = \alpha n^\alpha + \beta^\alpha, \qquad \beta^\alpha n_\alpha = 0.
\label{eq:ka}\eeq
Then, in a chart $\{t,x^i\}$, we have $t^\alpha = \partial_t$, and the metric
$g_{\alpha\beta} =  \gamma_{\alpha\beta} - n_\alpha n_\beta$ has the form
\beq
\label{eq:metric}
ds^2 = -\alpha^2 dt^2 +\gamma_{ij}(dx^i+\beta^i dt)(dx^j+\beta^j dt).
\label{gmunu}
\eeq
With a spatial covariant derivative $D_a$ compatible with 
the spatial metric $\gamma_{ab}$, the extrinsic curvature of 
$\Sigma_t$ is given by
\beq
K_{ab}  = -\frac12\Lie_n \gamma_{ab}
                =\frac1{2\alpha} (-\pa_t \gamma_{ab}
                        +D_a \beta_b + D_b \beta_a),
\label{eq:kab}
\eeq
where $\gamma_{ab}$ and $\pa_t\gamma_{ab}$ are the pullbacks to $\Sigma_t$ 
of $\gamma_{\alpha\beta}$ and $\Lie_t\gamma_{\alpha\beta}$.

In the canonical formulation of general relativity\cite{MTW},
$\{\gamma_{ab}, \pi^{ab},\alpha$ and $\beta^a\}$  
are regarded as independent gravitational field variables, 
where $\pi^{ab}$ is defined by 
\beq
\pi^{ab} := - (K^{ab}-\gamma^{ab} K)\sqrt{\gamma}. \label{piab}
\eeq
A perfect fluid is described by a stress-energy tensor 
\beq
\Tabu=(\epsilon+p)u^\alpha u^\beta + p \gabu, 
\eeq
where $u^\alpha$, $p$, and $\epsilon$ are the fluid four-velocity, 
pressure, and energy density, respectively.  The pressure and the 
energy density are assumed to satisfy an equation of state of the form
\beq
p=p(\rho,s), \quad \epsilon=\epsilon(\rho,s),
\eeq
where $\rho$ is the baryon mass density and $s$ the entropy per unit 
baryon mass.  

In calculating the variation of the Lagrangian 
following Routh procedure, 
a perfect-fluid spacetime is specified by the canonical variables,  
the lapse and the shift, that together describe the metric, and by 
Lagrangian variables for the fluid,  
$Q(\lambda):=[\gmabd(\lambda),\piabu(\lambda),\alpha(\lambda),
\beta^a(\lambda),u^\alpha(\lambda),\rho(\lambda),s(\lambda)]$.  
The difference between two nearby solutions can be treated 
in either of two ways.  Changes in the metric variables
will be written as Eulerian variations, denoted by $\delta$; the Eulerian 
change is the difference between corresponding quantities in the two solutions 
at a fixed point in spacetime.  Changes in fluid variables will be 
written as Lagrangian variations.  Introducing a Lagrangian displacement 
vector field $\xi^{\alpha}$, one defines the Lagrangian change in any 
fluid variable as the change with respect to a frame dragged by $\xi^{\alpha}$. Formally
The Lagrangian change $\Delta Q$ in a quantity $Q$ is then related to the 
Eulerian change $\delta Q$ by 
\beq
\Delta Q = \delta Q + \Lie_{\xi} Q.
\eeq
The description of fluid perturbation in terms of
a Lagrangian displacement $\xi^{\alpha}$ has a gauge freedom associated 
with a class of trivial displacements that yield no 
Eulerian change in the fluid variables.  We use this freedom 
to choose a gauge in which $\xi^t:=\xi^\alpha\na_\alpha t=0$, 
following \cite{F,FUS}.

The Einstein-Hilbert action 
\beq
S=\int \Lag d^4x, 
\label{eq:action} 
\eeq
with the Lagrangian density 
\beq
\Lag=\left(\frac1{16\pi}{}^4R-\epsilon\right)\sqrt{-g}, 
\label{eq:lagden}
\eeq
takes, in terms of Hamiltonian metric variables, the form 
\beq
16\pi{\cal L}
=  \pi^{ab}\pa_t\gamma_{ab}
        - \alpha\HG - \beta_a \CG^a
        + D_a (-2D^a\alpha\sqrt{\gamma} - 2\beta^b\pi^a{}_b + \beta^a\pi)
        -\pa_t\pi 
        -16\pi\,\epsilon\sqrt{-g},
\label{eq:lg}
\eeq
where ${}^4R$ is the Ricci scalar, 
\beq
\HG  := - 2 \Gabu n_\alpha n_\beta\,\sqrt{\gamma}\,, ~~~
\CG^a := -2 \Gabu \gamma_\alpha{}\!^a n_\beta\,\sqrt{\gamma}\, , 
\label{eq:HGCG}
\eeq
and $G^{\alpha\beta}$ is the Einstein tensor. 

The variation in the Lagrangian density is given by 
\beqn
\delta {\cal L} &=& - \rho T \sqrt{-g}\Dl s
            - \frac{h}{u^t}\Dl (\rho u^t\, \sqrt{-g})
\nonumber\\
      &&+\frac1{16\pi}\left[
        -\delta\alpha {\cal H} - \delta\beta^a {\cal C}_a
        + \delta \pi^{ab} \left\{\pa_t\gamma_{ab}-D_a\beta_b - D_b\beta_a
        -\frac{2\alpha}{\sqrt{\gamma}}
         \left(\pi_{ab} -\frac12\gamma_{ab}\pi\right)\right\}
        -\delta\gamma_{ab}(G^{ab}-8\pi S^{ab})\alpha\sqrt{\gamma}\right]
        \nonumber\\
&&- \xi_\alpha\na_\beta\Tabu\sqrt{-g}
  + D_a\tilde\Theta^a \sqrt{\gamma}
 - \frac1{16\pi}\pa_t(\delta \pi^{ab}\, \gamma_{ab})
 + \pa_t(j_a \xi^a \sqrt{\gamma}) \ ,
\label{eq:lham}
\eeqn
where $T$ is the temperature,
$h$ is the enthalpy defined by $h:=(\epsilon+p)/\rho$, and 
\beq
G^{ab} := \Gabu\gamma_\alpha\!{}^a\gamma_\beta\!{}^b  \ \mbox{ and }\ 
S^{ab} := \Tabu\gamma_\alpha\!{}^a\gamma_\beta\!{}^b. 
\eeq
With definitions
\beq
\rhoH:=\Tabu n_\alpha n_\beta~~~{\rm and}~~~
j^a:=-\Tabu \gamma_\alpha{}\!^a n_\beta,
\eeq
we set 
\beqn
{\cal H}  &:=& - 2 (\Gabu-8\pi\Tabu)n_\alpha n_\beta\,\sqrt{\gamma}
            \,=\,\HG+16\pi\,\rhoH
            \,=\,- \ \Bigl[ R
               - \frac1{\gamma}(\pi_{ab}\pi^{ab} - \frac12\pi^2)
               - 16\pi\rhoH\,\Bigr]\sqrt\gamma,
\label{eq:ham}\\
 {\cal C}^a &:=& -2(\Gabu-8\pi\Tabu)\gamma_\alpha{}\!^a n_\beta\,\sqrt{\gamma}
              \,=\,\CG^a-16\pi\, j^a\sqrt{\gamma}
              \,=\,-2(D_b\piabu + 8\pi j^a \sqrt{\gamma}),
\label{eq:mom}
\eeqn
where $R$ is the Ricci scalar with respect to $\gamma_{ab}$. 
The density $\tilde\Theta^a$ is the surface term of the Lagrangian 
density, 
\beqn
\tilde\Theta^a &=& \frac{1}{16\pi} 
\left[\frac{1}{\sqrt{\gamma}}
\Big\{-2\delta (D^a\alpha\, \sqrt{\gamma})
+ (\beta^a\gamma_{bc}\delta\pi^{bc} + \pi\delta\beta^a
- 2\pi^a{}_b \delta\beta^b)\Big\}\right.
\nonumber\\
&& \left.\phantom{\frac12}
+(\gamma^{ac}\gamma^{bd} - \gamma^{ab}\gamma^{cd})
(\alpha D_b\delta\gamma_{cd} -D_b\alpha\,\delta\gamma_{cd})\right]
+ \alpha (\epsilon +p)q^a{}_b\xi^b - \beta^a j^b\xi_b ,
\label{eq:tildetheta}
\eeqn
where $q^{ab}:=(\gabu+u^\alpha u^\beta)
\gamma_\alpha\!{}^a\gamma_\beta\!{}^b$.  

Independently varying the metric variables, 
$\{\dl\alpha,\dl\beta^a,\dl\gmabd, \dl\piabu\}$, gives the field equations, 
\beq
{\cal H}=0,\quad {\cal C}_a=0,\ \ \mbox{and}\ \  G^{ab}-8\pi S^{ab}=0,
\label{eq:fieldeqs}
\eeq
and the relation, 
\beq
\pa_t\gamma_{ab}-D_a\beta_b - D_b\beta_a
        -\frac{2\alpha}{\sqrt{\gamma}}
         \left(\pi_{ab} -\frac12\gamma_{ab}\pi\right) = 0.
\label{eq:defpi}
\eeq
Equation (\ref{eq:defpi}) is consistent with 
the definition of $\piabu$ [cf. Eq. (\ref{piab})].  

 When the field equations are satisfied, the Bianchi identity implies 
$\na_\beta \Tabu=0$.  The variation of the action with 
respect to the (spatial) Lagrangian displacement vector is the spatial 
projection of this relation, the relativistic Euler equation, 
\beq
\gamma^a{}\!_\alpha \na_\beta \Tabu=0. 
\label{eq:euler}
\eeq
For an isentropic fluid,  
conservation of baryon mass and entropy are given by 
\beq
\Lie_u(\rho\sqrt{-g})=0 \ \ \mbox{and}\ \ \Lie_u s=0.
\label{eq:conserv}
\eeq
Equations (\ref{eq:conserv}) and (\ref{eq:euler}) together imply 
$\na_\beta \Tabu=0$.  


It is often convenient to rewrite the above set of 
basic equations in terms of the conformally related spatial metric 
$\tilde \gamma_{ab}$ and the tracefree part of the extrinsic curvature 
$\tilde A_{ab}$, defined by 
\beqn
&& \tilde \gamma_{ab}:=\psi^{-4}\gamma_{ab}, \label{eq:cft} \\
&& \tilde A_{ab}:=\psi^{-4} \biggl(K_{ab}-{1 \over 3} \gamma_{ab} K \biggr),
\label{eq:cft2}
\eeqn
where $\psi$ is a conformal factor and 
$K := K_{ab}\gamma^{ab}$.  Here, 
we may impose the condition, $\tilde \gamma :=
{\rm det}(\tilde \gamma_{ab})={\rm det}(\eta_{ab})
= :\eta$, where $\eta_{ab}$ is a flat 3-metric. 
In the following, indices of variables with a tilde, such as 
$\tilde A_{ab}$, $\tilde A^{ab}$, $\tilde \beta_a$, and 
$\tilde \beta^a(=\beta^a)$, are raised and lowered by 
$\tilde \gamma_{ab}$ and $\tilde \gamma^{ab}$, respectively. 

%
%

The Hamiltonian constraint ${\cal H}=0$ and the momentum constraint 
${\cal C}_a=0$ are written in terms of $K_{ab}$: 
\beqn
&& R - K_{ab}K^{ab} +K^2 = 16\pi \rhoH, \label{ham} \\
&& D_b K^b_{~a} - D_a K = 8\pi j_a . \label{mom}
\eeqn
With the conformal transformation (\ref{eq:cft}) and
(\ref{eq:cft2}), Eqs.~(\ref{ham}) and 
(\ref{mom}) are rewritten in the form 
\beqn
&& \tilde \Delta \psi = {\psi \over 8}\tilde R- 2\pi \rhoH \psi^5 
-{\psi^5 \over 8} \Bigl(\tilde A_{ab} \tilde A^{ab}
-{2 \over 3}K^2\Bigr), \label{psieq} \\
&& \tilde D_b (\psi^6  \tilde A^b_{~a}) - {2 \over 3} \psi^6 
\tilde D_a K = 8\pi j_a \psi^6. \label{momeqf}
\eeqn
Here, $\tilde R$, $\tilde D_a$, and $\tilde \Delta$ are 
the Ricci scalar, the covariant derivative, and 
the Laplacian with respect to $\tilde \gamma_{ab}$, respectively.  

The evolution equations for $\gamma_{ab}$ and $K_{ab}$ are 
\beqn
\pa_t \gamma_{ab} && =
-2\alpha K_{ab} + D_a \beta_b +  D_b \beta_a,\label{h0eq} \\
\pa_t K_{ab}
&&= \alpha R_{ab}
- D_a D_b \alpha  
+\alpha (K K_{ab} - 2 K_{ac} K_b^{~c}) 
~+(D_b \beta^c) K_{ca}+(D_a \beta^c) K_{cb}+(D_c K_{ab}) \beta^c
\nonumber \\
&&~~~~
-8\pi\alpha \biggl[ S_{ab}+{1 \over 2} \gamma_{ab}
\Bigl( \rhoH - S_c^{~c} \Bigr)\biggr], \label{k0eq}
\eeqn
where $R_{ab}$ is the Ricci tensor with respect to $\gamma_{ab}$. 

Contracting $\gamma^{ab}$ with Eqs. (\ref{h0eq}) and (\ref{k0eq}) 
and using Eq. (\ref{ham}), one obtains  
\beqn
\pa_t \psi && = {\psi \over 6}\biggl( -\alpha K + D_c \beta^c \biggr), 
\label{psieqe} \\
\pa_t K&&=\alpha K_{ab} K^{ab}
-\Delta \alpha +4\pi \alpha (\rhoH + S_a^{~a})+\beta^a \pa_a K,
 \label{ktreq}
\eeqn
where $\Delta = D_a D^a$. 
To write the evolution equation of $K$ 
in the form of Eq. (\ref{ktreq}), we use the Hamiltonian 
constraint equation (\ref{ham}). 

In the following, we choose the maximal
time slicing condition $K=0=\pa_t K$. With this condition,
Eq. (\ref{ktreq}) reduces to an elliptic equation for $\alpha$, 
\beq
\Delta \alpha =\alpha \tilde A_{ab} \tilde A^{ab} + 4\pi \alpha
(\rhoH + S_a^{~a}),
\label{alpeq0}
\eeq
where we keep spatial indices abstract 
until fixing the spatial gauge condition.
Using Eq. (\ref{psieq}), this equation is rewritten 
\beq
\tilde \Delta (\alpha \psi)
=2\pi \alpha \psi^5 (\rhoH + 2S_a^{~a})
+{7 \over 8}\alpha \psi^5 \tilde A_{ab} \tilde A^{ab}
+{\alpha \psi \over 8} \tilde R~. \label{alpeq}
\eeq
Using Eqs. (\ref{h0eq}), (\ref{k0eq}), (\ref{psieqe}), and (\ref{ktreq}), 
the evolution equations for $\tilde \gamma_{ab}$ and 
$\tilde A_{ab}$ are 
\beqn
&&\pa_t \tilde \gamma_{ab}
=-2\alpha \tilde A_{ab} 
+\tilde D_a \tilde \beta_b + \tilde D_b \tilde \beta_a
-{2 \over 3}\tilde \gamma_{ab} \tilde  D_c \tilde\beta^c , \label{heq} 
\\
&&\pa_t \tilde A_{ab} 
= \psi^{-4} \biggl[ \alpha \Bigl(R_{ab}
-{\psi^4 \over 3} \tilde \gamma_{ab} R \Bigr) 
-\Bigl( D_a D_b \alpha - {\psi^4 \over 3}
\tilde \gamma_{ab} D_c D^c \alpha \Bigr)
\biggr] \nonumber \\
&& \hskip 2.cm -2\alpha \tilde A_{ac} \tilde A_b^{~c} 
+\tilde D_a \beta^c \tilde A_{cb}+\tilde D_b \beta^c \tilde A_{ca}
-{2 \over 3} \tilde D_c \beta^c \tilde A_{ab} 
+\beta^c \tilde D_c \tilde A_{ab} \nonumber \\
&& \hskip 2.cm-8\pi\alpha \Bigl( 
\psi^{-4} S_{ab}-{1 \over 3} \tilde \gamma_{ab} S_c^{~c}
\Bigr). \label{aijeq} 
\eeqn

Now, $R_{ab}$ is split, 
\beq
R_{ab}=\tilde R_{ab}+R^{\psi}_{ab},
\eeq
where $\tilde R_{ab}$ is the Ricci tensor with respect to 
$\tilde \gamma_{ab}$ and 
\beqn
R^{\psi}_{ab}=-{2 \over \psi}\tilde D_a \tilde D_b \psi 
-{2 \over \psi} \tilde \gamma_{ab}\tilde \Delta \psi 
+{6 \over \psi^2} \tilde D_a \psi \tilde D_b \psi 
- {2\over \psi^2} \tilde \gamma_{ab} \tilde D_c \psi \tilde D^c \psi. 
\eeqn
$\tilde R_{ab}$ is then written in the form 
\beqn
\tilde R_{ab}&=&{1 \over 2}\biggl[
-\tilde \gamma^{cd}\zeroD_c \zeroD_d  \tilde \gamma_{ab}
-\zeroD_b (\tilde \gamma_{ac} F^c)
-\zeroD_a (\tilde \gamma_{bc} F^c) \nonumber \\
&&~~~
-(\zeroD_c \tilde \gamma_{bd}) \zeroD_a \tilde\gamma^{cd} 
-(\zeroD_c \tilde \gamma_{ad}) \zeroD_b \tilde\gamma^{cd} 
+2 F^c C_{c,ab}-2 C^d_{cb}C^c_{ad}\biggr] \nonumber \\
&=&{1 \over 2}\biggl[
-\zeroDelta  \thh_{ab}
-\zeroD_b (\tilde \gamma_{ac} F^c)
-\zeroD_a (\tilde \gamma_{bc} F^c)\biggr]
+\tilde R_{ab}^{\rm NL},\label{eqij}
\eeqn
where $F^a := \zeroD_b \tilde \gamma^{ab}$,
$\zeroD_a$ is the covariant derivative associated with $\eta_{ab}$,
and
$\zeroDelta =\eta^{cd}\zeroD_c \zeroD_d$.
$\tilde R_{ab}^{\rm NL}$ is the collection of the nonlinear terms
in $\thh_{ab}$ and defined by
\beq
\tilde R_{ab}^{\rm NL}:=
-{1 \over 2}\biggl[ \tf^{cd}\zeroD_c \zeroD_d \thh_{ab}
+(\zeroD_c \thh_{bd}) \zeroD_a \tf^{cd} 
+(\zeroD_c \thh_{ad}) \zeroD_b \tf^{cd} \biggr]
+F^c C_{c,ab}-C^d_{cb}C^c_{ad}. \nonumber \\
\eeq
Here, $\thh_{ab}$ and $\tf^{ab}$ are introduced, respectively, by 
\beq
\thh_{ab}:=\tgmabd-\eta_{ab}\ \  \mbox{and}\ \ \tf^{ab}:=\tgmabu-\eta^{ab}. 
\label{eq:defhab}
\eeq
$C^c_{ab}$ and $C_{c,ab}$ are defined by 
\beqn
C^c_{ab} := 
{\tilde \gamma^{cd} \over 2}\Bigl(\zeroD_a \thh_{bd}
+\zeroD_b \thh_{ad}-\zeroD_d \thh_{ab}\Bigr) ~~{\rm and}~~
C_{d,ab} := \tilde \gamma_{cd} C^c_{ab},
\eeqn
and $C^c_{cd}=\zeroD_d(\sqrt{\tilde \gamma/\eta})/\sqrt{\tilde \gamma/\eta}=0$, 
when $\tilde \gamma=\eta$. 

\subsection{Basic equations for quasiequilibria}
\label{sec:basiceqs}

Compact binary systems in quasiequilibrium circular orbits 
evolve toward merger due to gravitational radiation reaction. 
Since the emission time scale of gravitational waves is longer
than the orbital period even just before the merger,
we may expect that the fluid and field variables near the support of 
fluid source (or inside the light cylinder) in the frame rotating 
with the same angular velocity as the orbital motion 
are approximately unchanged along a direction of a helical vector 
\beq
k^\alpha=t^\alpha + \Omega \phi^\alpha, 
\eeq
where $\phi^\alpha$ is a spatial vector field which generates 
a family of closed circular curves on $\Sigma_t$, and $\Omega$ 
denotes the orbital angular velocity. 

First, we derive hydrodynamic equations to describe binary neutron stars
in quasiequilibrium circular orbits.  
The baryon mass conservation law Eq.~(\ref{eq:conserv}) 
and the Euler equation Eq.~(\ref{eq:euler}) are written 
\beqn
&& \Lie_{k+v}(\rho u^t \sqrt{-g})=0,\label{eq:conti2}\\
&& \gamma_a\!^{\alpha}\Lie_{k+v}(h u_\alpha) \,
+\, D_a \left(\frac{h}{u^t}\right)=0.
\label{eq:euler2}
\eeqn
Then, we impose the conditions that the Lie derivatives along $k^\alpha$ 
vanish 
\beqn
&& \Lie_{k}(\rho u^t \sqrt{-g})=0,
\label{eq:heliden} \\
&& \gamma_a\!^{\alpha}\Lie_{k} (h u_\alpha)=0.
\label{eq:helivel}
\eeqn
Here, a spatial velocity vector $v^\alpha$ is introduced by 
\beq
u^\alpha = u^t(k^\alpha+v^\alpha). 
\label{eq:defvel}
\eeq
From conditions (\ref{eq:heliden}) and (\ref{eq:helivel}), 
the relation $\Lie_{k} (j_a \sqrt{\gamma})=0$ also follows. 
In the above, we assumed isentropic flow, 
which leads to the local first law of the thermodynamics, 
\beq
{1 \over \rho} \nabla_\alpha p= \nabla_\alpha h.
\label{eq:thermo}
\eeq
Equation (\ref{eq:thermo}) also implies that a one-parameter equation 
of state may be chosen.  
We thus have four independent variables for the fluid, 
three for the fluid velocity and one thermodynamic variable, 
governed by four equations (\ref{eq:conti2}) and (\ref{eq:euler2}).  

%

For corotational flow $u^\alpha = u^t k^\alpha$ (that is $v^\alpha=0$) or 
irrotational flow $hu_\alpha = \na_\alpha \Phi$, where $\Phi$ is a velocity 
potential, one can 
obtain a first integral of the Euler equation that is useful for 
computing quasiequilibrium configurations.  
For corotational flow, the velocity field becomes trivial and 
the first integral is the statement that the injection energy 
is constant in the fluid:    
\beq
\frac{h}{u^t}=\mbox{constant}.
\eeq
For irrotational flow, one thermodynamic variable and 
a velocity potential are governed by two equations \cite{S98}, 
\beqn
&&D_a \left[\frac{\alpha\rho}{h}\left(D^a\Phi-h u^t\omega^a\right)\right]=0,\\
&&\Lie_v\Phi \,+\, \frac{h}{u^t}=\mbox{constant},
\eeqn
derived from Eqs. (\ref{eq:conti2}) and (\ref{eq:euler2}), 
respectively.  
Here, the Lie derivative with respect to the spatial vector $v^a$, 
$\Lie_v$, is defined on $\Sigma_t$ with a relation 
\beq
\dis v^a=\gamma^a\!_{\alpha}\left(\frac{u^\alpha}{u^t}-\omega^\alpha\right)
=\frac1{h u^t}D^a\Phi-\omega^a,
\eeq
where $\omega^\alpha$ is a rotational 
shift vector defined by $\omega^\alpha=\beta^\alpha+\Omega\phi^\alpha$. 

%
%

Note that the symmetry of the spacetime with respect to the helical 
vector $k^\alpha$ has not been imposed yet. 
Thus, the Lie derivatives of the fluid quantities along $k^\alpha$
may not vanish, e.g., $\Lie_{k} \rho\neq 0$, $\Lie_{k} u_\alpha\neq 0$,
$\Lie_{k} \rhoH\neq 0$, and $\Lie_{k} S_{ab}\neq 0$.  
The values of these quantities depend 
on the formulation for gravitational fields that we choose below;
their magnitude measures the deviation from the helical symmetry.

We turn to the formulation for the gravitational fields of 
binary systems in quasicircular orbits. 
%
%
Here, we do not assume a global helical symmetry for the 
whole spacetime. 
First, we define $u^{ab} := \pa_t \tilde \gamma^{ab}$ and regard it as
an input quantity: 
It is determined when we impose a certain condition
between $\tilde \gamma^{ab}$ on two spatial hypersurfaces of 
infinitesimal time difference, following the concept of a thin
sandwich formalism proposed by York \cite{York}.  
In this section, we continue calculation 
without fixing the condition for $u^{ab}$ except for a requirement 
\beq
u^{ab}=\pa_t \tilde \gamma^{ab}=O(r^{-3}),  \label{uconst}
\eeq
in a far zone ($r \gg 2\pi\Omega^{-1}$).  
This condition guarantees the asymptotic flatness of the system on 
a slice $\Sigma_t$, but 
breaks the helical symmetry for $\tilde \gamma^{ab}$ in the far zone.  
The lower component of the time derivative is defined by
\beq
u_{ab} := -\tilde\gamma_{ac}\tilde\gamma_{bd}u^{cd}. \label{udown}
\eeq
We also define 
\beq 
v_{ab}:=\pa_t \hat A_{ab},
\eeq
where $\hat A_{ab} := \psi^6 \tilde A_{ab}$. Later, we will 
impose a condition on $v_{ab}$. 

Bonazzola et al. \cite{BGGN04} propose a ``gravito-inelastic 
approximation'' in which they set $u^{ab} =0$,
while $\pa_t \hat A^{ab}$ is free.  On the other hand, 
Sch\"afer and Gopakumar \cite{SG04} propose a minimal 
no-radiation approximation, in which the transverse-tracefree part of 
$\pa_t \piabu$ is restricted. We determine the conditions 
in the more rigid way: 
Our conditions for $u^{ab}$ and $v_{ab}$ are determined 
from the requirement that the quasiequilibrium solution
and its sequence satisfy the virial relation
and first law at least approximately. 
This subject will be discussed in Secs.~III and IV. 

Equation (\ref{heq}) is regarded as the equation for
determining $\tilde A_{ab}$, namely, 
\beq
2\alpha \tilde A_{ab} = 
\tilde D_a \tilde \beta_b + \tilde D_b \tilde \beta_a
-{2 \over 3}\tilde \gamma_{ab} \tilde  D_c \tilde\beta^c
-u_{ab}. \label{aaabb}
\eeq
Then, substituting Eq. (\ref{aaabb}) into Eq. (\ref{momeqf}), we obtain 
\beq
\tilde \Delta \tilde \beta_a 
+{1 \over 3}\tilde  D_a \tilde D_b \tilde \beta^b +
\tilde R_{ab} \tilde \beta^b 
+\tilde D^b \ln\biggl({\psi^6 \over \alpha}\biggr)
\Bigl[\tilde D_b \tilde \beta_a + \tilde D_a \tilde \beta_b - 
{2 \over 3}\tilde\gamma_{ab}\tilde D_c \tilde \beta^c \Bigr]
-{\alpha \over \psi^6}
\tilde D_c \Big(\alpha^{-1} \psi^6 \tilde\gamma^{bc}
u_{ab} \Big) =16\pi \alpha j_a .
\label{betaeq}
\eeq
This elliptic equation determines $\tilde\beta_a$. 

Regarding $v_{ab}$ as an input quantity, the 
evolution equation for $\tilde A_{ab}$ may be rewritten as 
an elliptic equation for $\thh_{ab}$, 
\beqn
&&\zeroDelta \thh_{ab}
=  2\biggl[ R^{\psi}_{ab}
-{1 \over 2}\zeroD_a (F^c \tilde \gamma_{cb})
-{1 \over 2}\zeroD_b (F^c \tilde \gamma_{ca})  + R^{\rm NL}_{ab}
-{\psi^4 \over 3} \tilde \gamma_{ab} R 
-{1 \over \alpha} \Bigl( D_a D_b \alpha - {\psi^4 \over 3}
\tilde \gamma_{ab} \Delta \alpha \Bigr)\biggr] \nonumber \\
&& \hskip 2.5cm -4 \psi^4 \tilde A_{ac} \tilde A_b^{~c} 
+{2 \psi^4 \over \alpha} \biggl(
\tilde D_a \beta^c \tilde A_{cb}+\tilde D_b \beta^c \tilde A_{ac}
-{2 \over 3} \tilde D_c \beta^c \tilde A_{ab} 
+\beta^c \tilde D_c \tilde A_{ab} \biggr)\nonumber \\
&& \hskip 2.5cm -16\pi \Bigl( 
S_{ab}-{\psi^4 \over 3} \tilde \gamma_{ab} S_c^{~c}\Bigr) 
-{2\psi^4\over \alpha}\pa_t \tilde A_{ab}.
\label{hijeq}
\eeqn
Together with the above equations for $\thh_{ab}$ and $\tilde\beta_a$, 
the elliptic equations (\ref{psieq}) and (\ref{alpeq}) 
are solved for $\psi$ and $\chi := \alpha\psi$, respectively.  

To summarize, the Einstein equations are rewritten as 
four elliptic equations (\ref{psieq}), (\ref{alpeq}), 
(\ref{betaeq}), and (\ref{hijeq}) for $\psi$, $\chi$, $\beta^a$, and 
$\thh_{ab}$, with $u^{ab}=\pa_t\tilde\gamma^{ab}$ and
$v_{ab}=\pa_t \hat A_{ab}$
which are regarded as input quantities that satisfy
a certain ansatz.  The asymptotic behavior of $u^{ab}$ and $v_{ab}$
is chosen to preclude standing waves in the far zone, 
$r \gg 2\pi \Omega^{-1}$. 
The simultaneous equations for the metric and the fluid on a slice 
$\Sigma_t$ is then similar to the equations for the initial value problem 
since the equations for the metric variables are elliptic. 

To solve the equations for $\thh_{ab}$, we need to fix the spatial gauge. 
The simplest choice is a transverse gauge \cite{SU01} (or a generalized
Dirac gauge in the terminology of \cite{BGGN04}), satisfying
\beq
F^a=\zeroD_b \tilde \gamma^{ab}=0. \label{TT}
\eeq
With this choice, the equations for $\thh_{ij}$ and
$\tilde R$ are significantly simplified, resulting in that 
the operator for the linear terms of $\thh_{ij}$ in its elliptic 
equation becomes the flat Laplacian. 
Furthermore, the behavior of the source terms of the elliptic
equations for $\psi$ and $\chi$
for $r \gg 2\pi\Omega^{-1}$ have suitable asymptotic behavior,
because  
$F^i  = O(r^{-3})$ and, hence, 
$\tilde R = O(r^{-4})$. Thus, 
the source terms of elliptic equations for $\psi$ and  
$\chi$ are $O(r^{-4})$ in the present gauge.
This implies that we can numerically solve 
these equations without serious difficulties. 

As a result of the present choice of gauge conditions,
the asymptotic behavior as $r \rightarrow \infty$ of the geometric
variables is \cite{OMur}
\beqn
&&\psi = 1 + {M_{\rm ADM} \over 2r} + O(r^{-2}), \label{bound1}\\
&&\alpha = 1 - {\MK \over r} + O(r^{-2}), \label{bound2}\\
&&\beta^k =
-{1 \over 4r^2}
\biggl(6 Z_{kl} \hat r^l + 3Z_{ij} \hat r^i \hat r^j \hat r^k
-Z_{ll} \hat r^k
+8Z^{\rm AS}_{kl} \hat r^l \biggr) + O(r^{-3}), \label{bound3}\\
&&\thh_{ij} = O(r^{-1}), \quad
	\partial_k\thh_{ij} = O(r^{-2}),\quad
	\partial_k\partial_l\thh_{ij} = O(r^{-3}),\label{bound4}\\
&&\tilde A_{ij} = 
-{1 \over 4r^3}
\biggl[ 6 Z_{ij}-2\delta _{ij}Z_{kk}
-6 Z_{il} \hat r^l \hat r^j 
-6 Z_{jl} \hat r^l \hat r^i
+4 \delta_{ij} Z_{kl} \hat r^k \hat r^l \nonumber \\
&&~~~~~~~~~~
+(5Z_{kl} \hat r^k \hat r^l -Z_{ll})(\delta_{ij}-3 \hat r^i \hat r^j)
-12(Z^{\rm AS}_{ik} \hat r^k \hat r^j
+Z^{\rm AS}_{jk} \hat r^k \hat r^i) + U_{ij}
\biggr] + O(r^{-4}),\label{bound6} \label{asym}
\eeqn
where $M_{\rm ADM}$ and $\MK$ denote the 
ADM mass and Komar mass (see Sec. III), $\hat r^k=x^k/r$, and 
$Z_{kl}$ and $Z^{\rm AS}_{kl}$ are time-dependent symmetric
and antisymmetric moments, respectively.
Equation (\ref{bound6}) implies that the total linear momentum
of the system is implicitly assumed to be zero, as 
\beq
\int_\infty \tilde A_j^{~i}dS_i=
\int_\infty \psi^6 \tilde A_j^{~i}dS_i=0. 
\eeq
Here
\[ \int_\infty :=\lim_{r\rightarrow\infty}\int_{S_r}\ ,  \]
with $S_r$ a sphere of constant $r$.  
The antisymmetric moment can be related to the angular momentum
of the system as
\beq
Z^{\rm AS}_{kl}=-J_j \epsilon_{jkl},
\eeq
where $\epsilon_{jkl}$ is the completely antisymmetric symbol. 
In the Newtonian limit,
\beqn
&&Z_{kl} =  \int \rho (v^k x^l + v^l x^k)
d^3x={d I_{kl} \over dt},\\
&&Z^{\rm AS}_{kl} = \int \rho (v^k x^l -v^l x^k)d^3x, 
\eeqn
where $I_{ij}$ is the quadrupole moment. 
$U_{ij}$ in Eq. (\ref{asym}) is a symmetric tracefree moment determined
by the asymptotic behavior of $u_{ij}$.

A solution to the simultaneous equations derived here satisfies the 
constraint equations of the Einstein equation. 
In this sense, it can be referred to as a fully general
relativistic solution and can be used as an initial condition of the 
(3+1) numerical simulation. However, 
since we do not assume the helical symmetric relation 
$\Lie_{k} \gamma_{ab}\not=0$, the solution does not 
satisfy the equation for quasiequilibrium exactly. 
As a result, $\Lie_{k} \psi$, $\Lie_{k} \alpha$,
and $\Lie_{k} \beta^a$ would be slightly different from zero in general. 
Deviation of these quantities from zero can measure the
violation of the helical symmetry. 
The magnitude of the deviation depends on our choice of
$u^{ab}$ and $v_{ab}$.

\section{Relations between $\Madm$ and $\MK$}
\label{sec:Madm=MK}

In this section, we derive the conditions needed for 
equality of the ADM mass $\Madm$ and the Komar mass $\MK$.
This equality is closely related to 
the virial relations as discussed in Appendix A. 

\subsubsection{Sufficient conditions for equality of $\MK$ and $\Madm$}

The Komar mass\cite{komar59} is constructed 
from a vector $\zeta^\alpha$ that approaches a timelike Killing vector of a 
flat asymptotic metric at spatial infinity.  As presented in 
\cite{komar62}, $\zeta^\alpha  = -\alpha^2\nabla^\alpha t$, 
and     
\beq 
\MK = {1 \over 8\pi} \int_\infty 
(\nabla^\beta \zeta^\alpha-\nabla^\alpha \zeta^\beta)n_\alpha dS_\beta.
\label{mk1}
\eeq 
With the metric written in the form (\ref{gmunu}), one uses the 
relations $n_\alpha = \alpha\nabla_\alpha t$ and 
$n^\beta\nabla_\beta n^\alpha = \alpha^{-1} \gamma_\alpha^a D_a\alpha$,
to obtain
\beqn
\nabla_\beta(\nabla^\beta \zeta^\alpha-\nabla^\alpha \zeta^\beta)n_\alpha
= \alpha \nabla_\beta\{\gamma^\beta_\gamma [\nabla^\alpha(\alpha n^\gamma)
			-\nabla^\gamma(\alpha n^\alpha)]\nabla_\alpha t\} 
= 2 D_a D^a \alpha. \label{eq3.2}
\eeqn
Using Eq. (\ref{eq3.2}), we have 
\beq
\frac1{8\pi}\int 2 D_a D^a \alpha\, dV
	= \frac1{4\pi}\int_\infty D^a\alpha\ dS_a, 
\eeq
and thus, we reach the familiar form \cite{misner} 
\beq 
\MK = \frac1{4\pi}\int_\infty D^a\alpha\ dS_a.
\label{mk2}\eeq

For a stationary spacetime, with $\zeta^\alpha$ the asymptotically 
timelike Killing vector, Beig \cite{beig} and Ashtekar and 
Ashtekar \cite{ashtekar} prove the equality $\MK =\Madm$. We 
obtain here more general asymptotic conditions sufficient for equality 
in the following way (patterned in part on Beig's work). 
Suppose that the metric has the form (\ref{gmunu}), 
with\footnote{The definition $h_{ij}$ here is slightly different from 
$\thh_{ij}$ in Eq.~(\ref{eq:defhab}) in Sec.~\ref{sec:basiceqs}.} 
\beq
 \gamma_{ij} = \eta_{ij} + h_{ij}, 
\label{asymp1}
\eeq
\beq
h_{ij} = O(r^{-1}), \quad D_k h_{ij} = o(r^{-1}), 
\quad D_k D_l h_{ij}= o(r^{-3/2}), 
\eeq
and suppose that the lapse, shift, and extrinsic curvature satisfy
\beqn
\alpha &=& 1-\frac{\alpha^{\!\!\!\! 1} }{r}\;+\;o(r^{-1}),  \quad 
\partial_r \alpha = \frac{\alpha^{\!\!\!\! 1} }{r^2}+o(r^{-2}),  \quad
\partial_r^2\alpha = -2 \frac{\alpha^{\!\!\!\! 1} }{r^3}+o(r^{-3});
\label{asymp2}\\
\beta^i &=& O(r^{-1-\epsilon}), \quad \partial_j\beta^i = o(r^{-3/2});
\label{asymp3}\\
\int_\infty d\hat\Omega K_{rr} &=& o(r^{-3}),
\qquad K_{ij} = o(r^{-3/2}), \quad 
      \partial_k K_{ij} = o(r^{-2+\epsilon}),
\label{asymp4}
\eeqn
where $\int d\hat\Omega$ denotes a surface integral. 
Then $\MK = \Madm$. 

To prove this claim, it is useful to introduce $\delta\,{}^3G_{ij}$, 
the part of $^3G_{ij}$ linear in $h_{ij}$.  The idea is to show that 
if $\delta\,{}^3G_{ij}$ is the asymptotically dominant part of $^3G_{ij}$,
then
\beq
\int_\infty {}^3G^i_j x^j\;d S_i = -8\pi \Madm.
\eeq
One then uses 
the field equation for $^3G_{ij}$ (the dynamical equation for 
$K_{ij}$) to show that this integral can be written in the form 
(\ref{mk2}) of the Komar mass, when $K_{ij}$, $\Lie_\beta K_{ij}$, 
and $\partial_t K_{rr}$ fall off rapidly enough at spatial infinity. 
  
Formally, 
\beq
\delta\,{}^3G_{ij} 
	 =  \frac{1}{2}\Big(\zeroD_i \zeroD_k h_j^k +\zeroD_j \zeroD_k h_i^k
 	-\zeroDelta h_{ij}-\zeroD_i \zeroD_j h^k_k
        + \eta_{ij} \zeroDelta h^k_k
 	-\eta_{ij} \zeroD_k \zeroD_l h^{kl}\Big),
\eeq
where the index of $h_{ij}$ is raised by the flat metric $\eta^{ij}$.  
The asymptotic behavior of $^3 G_{ij}$ is given by 
\beq
{}^3G_{ij} -\delta ^3G_{ij} = o(r^{3}).
\eeq
This is because each term in $^3G_{ij}$ involves either  
$D_k D_l h_{ij}$ or $D_k h_{ij} D_l h_{mn}$; then terms quadratic or 
higher order in $h_{ij}$ fall off as rapidly
as either $h_{mn} D_k D_l h_{ij}$ or $D_k h_{ij} D_l h_{mn}$.  

{}From the linearized Bianchi identity,
$\zeroD\!{}^j\delta\, {}^3G_{ij}=0$, we have
\beqn
 \int_\infty\;^3G^i_{\ j}x^j dS_i 
 	= \int \; \zeroD\!{}^i(\delta\, {}^3G_{ij}\,x^j)\sqrt\eta d^3x
 	=  \int \;\delta\, {}^3G_{ij}\;\eta^{ij}\sqrt\eta d^3x,
\eeqn
where
\beqn
\eta^{ij}\delta\, {}^3G_{ij}
	= -\frac{1}{2}\zeroD_i \zeroD_j h^{ij}+\frac{1}{2}\zeroDelta h^k_k\;\;
	= \frac{1}{2}\zeroD_i(\zeroD\!{}^i h^k_k\;-\;\zeroD_j h^{ij}).
\eeqn
Then
\beq
 \int_\infty {}^3G^i_j x^j\;d S_i 
 = -\frac{1}{2}\,\int_\infty(\zeroD_j h^{ij}-\zeroD\!{}^i h^k_k)dS_i 
	=: -8\pi \Madm.
\label{eq:asyMadm}
\eeq
  
Next, from Eqs. (\ref{k0eq}) and (\ref{ham}), $^3G_{ij}$ has, outside the 
matter, the form 
\beqn
{}^3G_{ij} &=& R_{ij} - \frac12 \gamma_{ij} R \nonumber \\
 &=& \frac1\alpha D_i D_j \alpha
- \frac12 \gamma_{ij} \frac1\alpha \Delta\alpha    
    +\frac1\alpha \Bigl(\partial_t K_{ij}
-\frac12\gamma_{ij}\gamma^{kl}\partial_t K_{kl}\Bigr)
    +2K_{ik}K_j^k - K_{ij}K \nonumber \\
 && - \gamma_{ij}\Bigl(K_{kl} K^{kl} - \frac12 K^2\Bigr)  
    + \frac1\alpha\Lie_\beta K_{ij} 
    -\frac1{2\alpha} \gamma_{ij}\gamma^{kl}\Lie_\beta K_{kl}.
\eeqn
Since the terms involving $K_{ij}$ are $o(r^{-3})$, we have  
\beq
\int_\infty\;^3G^i_j x^j\;dS_i 
	= \int_\infty D_i D_j\alpha\;\hat{r}^i\hat{r}^j r^3 d\hat\Omega. 
\eeq
Now $\partial_k\gamma_{ij}=O(r^{-2})$ and $\partial_i\alpha = O(r^{-2})$ 
imply
\beq
D_r^{\ 2}\alpha\;=\;\partial_r^{\ 2}\alpha\;+\;O(r^{-4}).
\eeq
Finally, from Eq.~(\ref{mk2}) and the asymptotic form (\ref{asymp2}) of 
$\alpha$, we have   
\beq
\MK = \frac1{4\pi}\int_\infty \alpha^{\!\!\!\! 1}\ d\hat\Omega,
\eeq
and 
\beq
\int_\infty\;^3G_j^ix^j\;dS_i=\int_\infty D_r^{\ 2}\alpha\;r^3\;d\hat\Omega 
	= -8\pi \MK,
\eeq
whence 
\beq 
\Madm = \MK.
\eeq
 
The asymptotic behavior of Eqs.~(\ref{bound1})--(\ref{asym}) does 
{\em not always} 
satisfy the conditions (\ref{asymp4}), because $\partial_t K_{ij}$ is 
$O(r^{-3})$, not $o(r^{-3})$.  


\subsubsection{Relation between $\MK$ and $\Madm$ 
in the waveless approximation}

For quasiequilibrium binary solutions 
that satisfy Eqs. (\ref{udown})--(\ref{hijeq}), with $K=0$, 
a slightly different relation holds between $\MK$ and $\Madm$.   
The asymptotic behavior given in Eqs.~(\ref{bound4}) and (\ref{bound1}) 
implies 
\beq
D_i h_{ij} = O(r^{-2}), \ \ \ \ D_i D_j h_{ij} = O(r^{-3}). 
\label{eq:dhij}
\eeq
For the asymptotic behavior of Eqs.~(\ref{bound1})--(\ref{bound6}), 
together with Eq.~(\ref{eq:dhij}), 
Eq.~(\ref{eq:asyMadm}) still holds and 
\beq
\int _\infty {}^3G^i_j x^j dS_i
  = -8\pi \MK + \int_\infty \partial_t K_{rr}r^3 d\hat\Omega
  = -8\pi \MK + \frac14\int_\infty \partial_t(12Z_{rr}-U_{rr}) d\hat\Omega,
\label{eq:virwat}
\eeq
or
\beq
\Madm  =  \MK -\frac{d}{dt} \left(\frac32 Z_{rr}-\frac18 U_{rr}\right)\ .
\eeq
Using 
\beq
\int_\infty \hat r^i \hat r^j d\hat\Omega = {4\pi \over 3}\delta_{ij},
\eeq
Equation (\ref{eq:virwat}) may be written in asymptotically Cartesian 
coordinates as 
\beq
\int _\infty {}^3G^i_j x^j dS_i
= -8\pi \MK + \int_\infty \partial_t K_{ij}\, \hat r^i\hat r^j r^3 d\hat\Omega
= -8\pi \MK + 4\pi \frac{dZ_{kk}}{dt}, 
\eeq
whence
\beq
\Madm  =  \MK - \frac1{2} \frac{dZ_{kk}}{dt}\ . 
\eeq
We note that a similar expression has been derived for 
the maximal slicing condition in
the first post Newtonian approximation \cite{ASF}.
The expression here is the fully general relativistic generalization.

In the presence of a timelike Killing vector, we may set 
\beqn
&&\pa_t \tilde\gamma_{ab}=0, \noindent \\
&&\pa_t (\psi^4 \pi^{ab})=0, \noindent \\
&&{d \over dt} Z_{kk}=0, \label{cond}
\eeqn
and, hence, the virial relation $\Madm=\MK$ holds. 
This condition should be satisfied not only for axisymmetric 
equilibria but also for nonaxisymmetric ones 
such as general relativistic Dedekind solutions.

The integral of $K_{ij}$ in the above expression may be computed further as
\beqn
\int_\infty \partial_t K_{ij}\, \hat r^i\hat r^j r^3 d\hat\Omega
&=& \frac1{8\pi}\frac{d}{dt} \int_\infty K_a{}^b x^a dS_b
= \frac1{8\pi}\frac{d}{dt} \int D_b (\piab x^a) d^3x \nonumber \\
&=& - \frac{d}{dt} \int x^a j_a dV + 
\frac1{16\pi}\frac{d}{dt} \int \piabu \,x^c\!\zeroD_c \gmabd d^3x, 
\label{eq:dKrr}
\eeqn
which yields 
\beqn
\Madm&=& \MK
 \,+\, \frac{d}{dt} \int x^a j_a\, dV
 \,-\, \frac1{16\pi}
\int (\pa_t \piabu\, x^c\! \zeroD_c \gmabd + \piabu\, x^c\! \zeroD_c \pa_t\gmabd) d^3x
\nonumber \\
&=& \MK
 \,+\, \frac{d}{dt} \int x^a j_a\, dV
 \,-\, \frac1{16\pi}
\int [\pa_t (\psi^4\piabu) \, x^c\! \zeroD_c \tgmabd 
+ \psi^4\piabu\, x^c\! \zeroD_c \pa_t\tgmabd] d^3x.
\label{eq:madmmk}
\eeqn

In a gauge with $K=0$, Eq.~(\ref{eq:madmmk}) can be written as 
\beqn
\Madm 
&=& \MK \,+\, \frac{d}{dt} \int x^a j_a\, dV
 \,-\, \frac1{16\pi}
\int [\pa_t \hat A_{ab}\, x^c\! \zeroD_c \tgmabu 
+ \hat A_{ab}\, x^c\! \zeroD_c \pa_t\tgmabu] d^3x 
\nonumber \\
&=& \MK \,+\, \frac{d}{dt} \int x^a j_a\, dV
 \,-\, \frac1{16\pi}
\int [v_{ab}\, x^c\! \zeroD_c \tgmabu 
+ \hat A_{ab}\, x^c\! \zeroD_c u^{ab}] d^3x,
\label{eq:madmmk2}
\eeqn
where a relation 
$\piabu \, x^c\!\zeroD_c\gmabd=\hat A_{ab}\, x^c\!\zeroD_c\tgmabu$ is used.  
Since we always impose a condition $\Lie_k (j_a\sqrt{\gamma})=0$, 
the second integral can be discarded because of the relation, 
\beq
{d \over dt} \int x^a j_a dV
=\int \pa_t (x^a j_a \sqrt{\gamma}) d^3 x
=\int \Lie_k (x^a j_a\sqrt{\gamma}) d^3 x= 0.
\eeq
Thus, Eq.~(\ref{eq:madmmk2}) is written as
\beq
\Madm =\MK \,-\,{1 \over 16\pi} \int
( v_{ab}\, x^c\! \zeroD_c \tilde \gamma^{ab} +\hat A_{ab}\, x^c\! \zeroD_c u^{ab}) d^3x. 
\label{simI2}
\eeq

If we assume that the spacetime is everywhere helically symmetric, $\Lie_k \gabd=0$, then it is not  
asymptotically flat 
and, hence, the integrals in Eqs.~(\ref{eq:madmmk}) and (\ref{eq:madmmk2}), 
as well as $M_{\rm ADM}$, are not defined. 
Instead, we require helical symmetry only in the 
near zone and the local distant zone.
In this case, in the local zone, we should set 
$\Lie_k\tgmabu=0$ and $\Lie_k \hat A_{ab}=0$ 
which can be written as 
\beqn
&&u^{ab}=\pa_t \tilde\gamma^{ab}
=-\Lie_{\Omega\phi} \tilde \gamma^{ab}, \noindent \\
&&v_{ab}=\pa_t \hat A_{ab}=-\Lie_{\Omega\phi} \hat A_{ab}, \label{cond2}
\eeqn
while in the distant zone, $u^{ab} \rightarrow 0$ and
$v_{ab} \rightarrow 0$ for $r\rightarrow \infty$. 
With these choices, $dZ_{kk}/dt$ vanishes, and hence, 
the virial relation $\Madm=\MK$ is satisfied. 
The virial relation is satisfied for the simple case 
$u^{ab}=v_{ab}=0$; Equation (\ref{simI2})
reminds us that it is satisfied in the conformal flatness approximation
$\tgmabu=\eta^{ab}$, i.e., $u^{ab}=0$ \cite{FUS}.


\section{Relation for $\delta \Madm$ and $\delta J$}
\label{sec:dM=omegadJ}

A first law for binary systems with a helical Killing field
$k^\alpha$ has been formulated as relating a change in a conserved
charge $Q$, associated with a family of helically symmetric spacetimes
\cite{FUS}, to the changes in the vorticity, baryon 
mass, and entropy of the fluid as well as 
in the area of black holes. In \cite{FUS}, we prove that 
a relation $\delta Q=\delta\Madm-\Omega\delta J$ 
holds for asymptotically flat systems, where $\Omega$ is the 
orbital angular velocity of a binary system in circular orbits. 

In inspiraling binary neutron stars, entropy, baryon 
mass, and vorticity are almost constant and, hence, 
energy and angular momentum are dissipated only by 
gravitational radiation. 
Thus, the relation $d\Madm/dt=\Omega dJ/dt$ is satisfied. Then, 
we should require that, 
in a formalism for computing asymptotically flat binary equilibria,
a first law $\delta\Madm=\Omega\delta J$ is satisfied. 
In this section, we present a heuristic way to derive a relation 
held between the variations of the ADM mass and the angular momentum 
without assuming any symmetry for perfect-fluid spacetimes, but 
requiring the field equations to be satisfied. Then, we 
identify sources for the violation of the first law.  
In the following calculation, no gauge condition is  specified; and surface terms associated with black hole horizons are not included. 

In contrast to our earlier paper \cite{FUS}, 
no helical symmetry is assumed.  
Instead, we assume only that the field equations derived
in Sec. II are satisfied. The field equations are used to
relate the Komar mass to the Lagrangian density and 
the other terms in the manner 
\beqn 
\MK&=&{1 \over 4\pi}\int_\infty D^a\alpha dS_a  
={1 \over 4\pi}\int D^a D_a\alpha dV \nonumber \\
&=& 2\int \left\{\rho u^t \sqrt{-g} h u_\alpha v^\alpha
-\Lag -\frac1{16\pi}\gamma_{ab}\pa_t\pi^{ab}
-\frac1{16\pi}D_a(2\pi^{ab}\beta_b -\pi\beta^a)\right\}d^3x,
\label{eq:kom}
\eeqn
where the definition of $v^\alpha$, slightly different from the 
previous section, is 
$u^\alpha=u^t(t^\alpha+v^\alpha)$ with $v^\alpha n_\alpha=0$.
To compute the last equality, we first use a part of
the Einstein equation, 
\beqn
&&\left(\Gabu - 8\pi\Tabu\right)
\left(\gamma_{\alpha\beta}+n_\alpha n_\beta
+\frac2{\alpha}\beta_\alpha n_\beta\right)\alpha\sqrt{\gamma}\nonumber \\
&=&\gamma_{ab}\pa_t\pi^{ab}
+2D^a D_a\alpha\sqrt{\gamma}
+16\pi\left(-\frac12 T_{\alpha}^{~\alpha}-\epsilon\right)\sqrt{-g}
-16\pi\rho u^t \sqrt{-g} h u_\alpha v^\alpha
+D_a(2\pi^{ab}\beta_b -\pi\beta^a)=0. 
\eeqn
We then subtract the trace of the Einstein equation 
$(\Gabu - 8\pi\Tabu)\gabd=-R-8\pi T_{\alpha}^{~\alpha}=0$
to relate $\MK$ to the Lagrangian density (\ref{eq:lagden}).

The angular momentum is defined by 
\beqn
J&=&-\frac1{8\pi}\int_\infty \pi^a{}_b\phi^b dS_a
=-\frac1{8\pi}\int D_a(\pi^a{}_b\phi^b) d^3x
=-\frac1{8\pi}\int (\phi^a D_b\pi^b{}_a+\pi^{ab}D_b\phi_a)d^3x 
\nonumber \\
&=&\int\left(j_a \phi^a \sqrt{\gamma}
-\frac1{16\pi}\piabu\Lie_\phi\gmabd\right)d^3x, 
\label{eq:angmom}
\eeqn
where the momentum constraint ${\cal C}^a=0$ is used in 
the last equality.

The variations of $\MK$ and $J$ are computed following \cite{FUS}.  
First, we take the variation of Eq.~(\ref{eq:kom}) for $\MK$, 
\beq
\dl \MK
= 2\int \left\{\Delta(\rho u^t \sqrt{-g} h u_\alpha v^\alpha)
-\dl\Lag -\frac1{16\pi}\dl(\gamma_{ab}\pa_t\pi^{ab})
-\frac1{16\pi}D_a\dl(2\pi^{ab}\beta_b -\pi\beta^a)\right\}d^3x, 
\eeq
and substitute the following relation into the first term, 
\beq
\Delta(\rho u^t \sqrt{-g} h u_\alpha v^\alpha)
= h u_\alpha v^\alpha \Delta(\rho u^t \sqrt{-g})
+\rho u^t \sqrt{-g} v^\alpha\Delta(h u_\alpha)
+ j_a \sqrt{\gamma} \pa_t \xi^a, 
\eeq
where we used the facts that a choice of $\xi^t=0$ implies 
$\Delta v^\alpha= -\Delta t^\alpha=\Lie_t \xi^\alpha$ and that 
$\Lie_t\xi^\alpha \na_\alpha t=0$ yields 
$\rho u^t \sqrt{-g} h u_\alpha \Lie_t \xi^\alpha=
j_a\sqrt{\gamma}\pa_t\xi^a$. 
To the second term, the variation in 
the Lagrangian density Eq.~(\ref{eq:lham}) 
is applied. 

When all components of the Einstein equation and the Bianchi identity, 
Eqs.~(\ref{eq:fieldeqs})--(\ref{eq:conserv}), are satisfied, 
$\dl \MK$ becomes
\beqn
\dl \MK &=& 2\int \left[
\rho T \Delta s \sqrt{-g}
+ \left(\frac{h}{u^t} + h u_\alpha v^\alpha\right)\Delta(\rho u^t \sqrt{-g})
+ v^\alpha\Delta(h u_\alpha) \rho u^t \sqrt{-g} 
- \xi^a\pa_t(j_a\sqrt{\gamma}) \right. \nonumber\\
&&+\left.\frac1{16\pi}(\dl \piabu \pa_t \gmabd - \dl \gmabd \pa_t \piabu)
- D_a\Big\{\tilde\Theta^a \sqrt{\gamma}
+ \frac1{16\pi}\dl(2\pi^{ab}\beta_b -\pi\beta^a)\Big\}\right] d^3x. 
\label{eq:dlkom2}
\eeqn
The contribution of the surface term of the above equation is given by 
\beqn
&-&2\int D_a\left[\tilde\Theta^a \sqrt{\gamma}
+ \frac1{16\pi}\dl(2\pi^{ab}\beta_b -\pi\beta^a)\right] d^3x 
=-2\int_\infty \tilde\Theta^a dS_a \nonumber\\
&=&\frac1{4\pi}\int_\infty D^a\dl\alpha dS_a
-\frac1{8\pi}\int_\infty
(\gamma^{ac}\gamma^{bd} - \gamma^{ab}\gamma^{cd})
D_b\delta\gamma_{cd} dS_a
=\dl \MK - 2\dl \Madm.
\label{eq:dlkom_surf}
\eeqn
Combining Eqs.~(\ref{eq:dlkom2}) and (\ref{eq:dlkom_surf}), 
the variation in the ADM mass $\Madm$, instead of Komar mass $\MK$, 
is written 
\beqn
\dl \Madm = \int \left\{
\rho T \Delta s \sqrt{-g}
+ \left(\frac{h}{u^t} + h u_\alpha v^\alpha\right)\Delta(\rho u^t \sqrt{-g})
\right.
&+& v^\alpha\Delta(h u_\alpha) \rho u^t \sqrt{-g} 
- \xi^a\pa_t(j_a\sqrt{\gamma}) \nonumber\\
&+&\left. \frac1{16\pi}(\dl \piabu \pa_t \gmabd - \dl \gmabd \pa_t \piabu)
\right\}d^3x \ .
\label{eq:dladm}
\eeqn

The variation in the angular momentum $\dl J$, computed in 
a similar way from Eq.~(\ref{eq:angmom}), is
\beqn
\dl J &=& \int\left\{\Dl(j_a\phi^a\sqrt{\gamma})
-\frac1{16\pi}\dl(\piabu\Lie_\phi\gmabd)\right\}d^3x \nonumber\\
&=&\int\left\{\phantom{\frac1{1}}\right.\!\!\!\!\!
h u_\alpha \phi^\alpha \Delta(\rho u^t \sqrt{-g})
+ \phi^\alpha\Dl(h u_\alpha) \rho u^t \sqrt{-g}
+ \xi^a \Lie_\phi(j_a \sqrt{\gamma})
\nonumber\\
&&\left.-\frac1{16\pi}(\dl\piabu\Lie_\phi\gmabd - \dl\gmabd\Lie_\phi\piabu)
-\Lie_\phi\left(j_a \xi^a \sqrt{\gamma}
+\frac1{16\pi} \piabu\dl\gmabd\right)\right\}d^3x, 
\label{eq:dlang}
\eeqn
where 
\beqn
\Dl(j_a\phi^a\sqrt{\gamma})
&=& \Delta(\rho u^t \sqrt{-g}) h u_\alpha \phi^\alpha
+\rho u^t \sqrt{-g} \phi^\alpha\Delta(h u_\alpha)
- j_a \sqrt{\gamma} \Lie_\phi \xi^a \nonumber\\
&=& \Delta(\rho u^t \sqrt{-g}) h u_\alpha \phi^\alpha
+\rho u^t \sqrt{-g} \phi^\alpha\Delta(h u_\alpha)
+ \xi^a \Lie_\phi (j_a \sqrt{\gamma})
- \Lie_\phi (j_a \xi^a \sqrt{\gamma}), 
\eeqn
which results from relations $j_a \phi^a \sqrt{\gamma}
=\rho u^t\sqrt{-g} h u_\alpha \phi^\alpha$, $\dl\phi^\alpha=0$, and   
$\Dl \phi^\alpha=-\Lie_\phi \xi^\alpha$.  
The last term in the integral of Eq.~(\ref{eq:dlang}) vanishes 
since it becomes a surface integral with a combination $\phi^a dS_a=0$. 

Finally, Eqs.~(\ref{eq:dladm}) and (\ref{eq:dlang}) 
are combined to derive a relation 
\beqn
\dl \Madm -\Omega \dl J = \int \biggl[
\rho T \Delta s \sqrt{-g}
&+&\left\{\frac{h}{u^t} + h u_\alpha (v^\alpha-\Omega \phi^\alpha)\right\}
\Delta(\rho u^t \sqrt{-g})
+ (v^\alpha-\Omega\phi^\alpha)\Delta(h u_\alpha) \rho u^t \sqrt{-g} 
\nonumber\\
&-& \xi^a\Lie_{t+\Omega\phi}(j_a\sqrt{\gamma}) 
+\frac1{16\pi}(\dl \piabu \Lie_{t+\Omega\phi} \gmabd
 - \dl \gmabd \Lie_{t+\Omega\phi} \piabu)
\biggr] d^3x , 
\label{eq:1stlaw}
\eeqn
where $\Lie_{t+\Omega\phi}=\pa_t+\Lie_{\Omega\phi}$, 
which operates on the spatial quantities, is understood as a pullback of 
the Lie derivative along the helical vector field onto $\Sigma_t$. 
The first three terms of Eq.~(\ref{eq:1stlaw}) are the same as those 
derived in our previous paper \cite{FUS} except for 
the definition of the spatial velocity 
$v^\alpha =u^\alpha/u^t$ (in \cite{FUS}, we used the definition 
$v^\alpha=u^\alpha/u^t-\Omega \phi^\alpha$). 
The difference also changes the definition of the shift.  

As discussed in \cite{FUS}, for an isentropic fluid, 
conservation of baryon mass, entropy, and vorticity become 
\beq
\Lie_u(\rho\sqrt{-g})=0, \quad \Lie_u s=0, \ \ \mbox{and}\ \ 
\Lie_u\omega_{\alpha\beta}=0.
\label{eq:conserv2}
\eeq
These imply perturbed conservation laws, 
\beq
\Dl(\rho u^t\sqrt{-g})=0, \quad \Dl s=0,  \ \ \mbox{and}\ \ 
\Dl\omega_{\alpha\beta}=0,
\label{eq:pconserv}
\eeq
that will be almost satisfied during binary inspiral before the merger. 
Here, the relativistic vorticity $\omega_{\alpha\beta}$ is given by
\beq
\omega_{\alpha\beta} =
 q_\alpha{}^\gamma q_\beta{}^\delta\left[\na_\gamma \left(h u_\delta\right)
          - \na_\delta \left(h u_\gamma\right)\right]
        = \na_\alpha \left(h u_\beta\right) -
        \na_\beta \left(h u_\alpha\right).
\eeq
The third term in Eq.~(\ref{eq:1stlaw}) vanishes for (i) corotating 
binaries, flows with $v^\alpha=\Omega\phi^\alpha$, and (ii) irrotational
binaries, potential flows with $hu_\alpha=\na_\alpha\Phi$.  For the latter 
case, the third term in Eq. (\ref{eq:1stlaw}) 
with $\Dl(hu_\alpha) = \Dl \na_\alpha\Phi = \na_\alpha \Dl\Phi$ becomes 
\beqn
&&\int(v^\alpha-\Omega\phi^\alpha)\Delta(h u_\alpha) \rho u^t \sqrt{-g}\,d^3x
=\int(v^\alpha-\Omega\phi^\alpha)\na_\alpha \Dl\Phi \rho u^t \sqrt{-g}\,d^3x
\nonumber\\
&&= \int \left[ D_\alpha\{(v^\alpha-\Omega\phi^\alpha) \Dl\Phi \rho u^t 
\alpha\sqrt{\gamma}\}
+\{\Lie_k(\rho\sqrt{-g}) - \Lie_u(\rho\sqrt{-g})\}\Dl\Phi \right]d^3x\
\nonumber \\
&&= \int [\Lie_k(\rho\sqrt{-g}) - \Lie_u(\rho\sqrt{-g})]\Dl\Phi d^3x,
\eeqn
where we assume $\rho=0$ for the distant zone to derive the last line. 
Then, together with Eqs.~(\ref{eq:conserv2}) and (\ref{eq:pconserv}),
Eq. (\ref{eq:1stlaw}) is rewritten as
\beq
\dl \Madm -\Omega \dl J = \int\left\{
\Lie_k(\rho\sqrt{-g})\Dl\Phi
- \xi^a\Lie_k(j_a\sqrt{\gamma}) 
+ \frac1{16\pi}(\dl \piabu \Lie_k \gmabd
 - \dl \gmabd \Lie_k \piabu)
\right\}d^3x . 
\label{eq:1stlaw2}
\eeq
The form of the ``first law'' described in Eq. (\ref{eq:1stlaw}) or 
Eq. (\ref{eq:1stlaw2}) is derived without relying on the 
helical symmetry of the spacetime and the fluid. 
Choosing the maximal slicing condition $K=0=\pi$, 
we may rewrite Eq.~(\ref{eq:1stlaw2}) 
using tracefree part of the conjugate momentum 
$\displaystyle \hat \piabu=\piabu-\frac{1}{3}\gmabu\pi$ 
and the conformal metric $\tgmabd =\psi^{-4}\gmabd$ as 
\beqn
\dl \Madm -\Omega \dl J &=& \int\left[
\Lie_k(\rho\sqrt{-g})\Dl\Phi
- \xi^a\Lie_k(j_a\sqrt{\gamma}) 
+ \frac1{16\pi}\{\dl (\psi^4 \hat\piabu) \Lie_k \tgmabd
- \dl \tgmabd \Lie_k (\psi^4 \hat\piabu)\}
\right]d^3x \nonumber \\ 
&=& \int\left[
\Lie_k(\rho\sqrt{-g})\Dl\Phi
- \xi^a\Lie_k(j_a\sqrt{\gamma}) 
+ \frac1{16\pi}(\dl \hat A_{ab} \Lie_k \tgmabu
- \dl \tgmabu \Lie_k \hat A_{ab})
\right]d^3x . 
\label{eq:1stlaw4}
\eeqn
Here, the gauge choice $\pi=0$ implies $\dl \pi=0$ 
and $\Lie_k \pi =0$.  

If one requires $\Lie_k \gmabd=0$ and $\Lie_k\piabu=0$ 
(or $\Lie_k\tgmabu=0$ and $\Lie_k \hat A_{ab}=0$) 
in a gauge $\pi=0=K$, together with conditions for the fluid variables
$\Lie_k(\rho\sqrt{-g})=0$ and $\Lie_k(j_a\sqrt{\gamma})=0$, 
Eq.~(\ref{eq:1stlaw2}) leads to the first law
relation $\dl \Madm =\Omega \dl J$.
However, these assumptions are equivalent to imposing  helical 
symmetry on the whole spacetime and, hence, preclude  
asymptotic flatness; in other words, $\Madm$ and $J$ are ill defined.  

In a realistic system, the radiation reaction violates 
the Killing symmetry.  
In its presence, $\delta \gmabd$ 
and $\delta \piabu$ are determined
by the radiation reaction, and these terms may be 
proportional to the violation of the helical symmetry 
near the source. 
Namely, we expect that the following relations hold:
\beqn
(\Lie_k \gmabd) \dl t = \dl \gmabd,\\
(\Lie_k \piabu) \dl t = \dl \piabu,
\eeqn
or in the gauge $K=0$,
\beqn
(\Lie_k \tgmabu) \dl t = \dl \tgmabu,\\
(\Lie_k \hat A_{ab}) \dl t = \dl \hat A_{ab},
\eeqn
where $\delta t$ is a radiation reaction time scale. 
In this case, the right-hand side of Eq.~(\ref{eq:1stlaw2}) vanishes
with the symmetry for the fluid variables.  
This indicates that even with the slight violation of the
helical symmetry due to the radiation reaction,
a relation $\dl \Madm =\Omega \dl J$ may be well satisfied. 

Finally, as shown in \cite{FUS}, $\dl \Madm =\Omega \dl J$ is 
exact in the conformal flatness approximation 
(IWM formalism).  In this case, one needs to replace the Lagrangian 
density (\ref{eq:lagden}) by one that reproduces the  
field equations of the IWM formalism.  One can derive such 
a Lagrangian density by substituting $\pi=0$ and $\tgmabd=\eta_{ab}$
into Eq.~(\ref{eq:lagden}).  Then, assuming helical symmetry 
for the fluid and from the fact $\delta \tgmabd=0$, 
the first law is shown to be satisfied.  
(See \cite{FUS} for a description of the artificiality of this
choice in a helically symmetric IWM framework.)

\section{Candidate formulations for quasiequilibria}
\label{sec:formalism}

The condition $u^{ab}=O(r^{-3})$ is not compatible with 
helical symmetry in the whole spacetime. 
Thus, we propose to impose 
\beq
u^{ab}=\left\{
\begin{array}{ll}
-\Lie_{\Omega \phi} \tilde \gamma^{ab}& {\rm for}~r \leq r_0, \\
0 & {\rm for}~r \geq r_0,
\end{array}
\right.
\eeq
where $r_0$ is an arbitrary radius.  With this condition,
the type of the field equation for $\tilde h_{ij}$ changes
from Helmholtz-type to elliptic for 
$r\simeq r_0$. To make the equation be almost elliptic 
for numerical computation, it may be desirable to take $r_0$
within the light cylinder radius as $r_0 \lesssim 2\pi/\Omega$.  
On the other hand, we can impose helical symmetry on $\hat A_{ab}$ 
without serious difficulty. In this case, helical symmetry is exact 
in the near zone and, 
as a result, the violation of the first law is given by 
\beqn
\delta \Madm - \Omega \delta J
&=&{1 \over 16\pi} \int_{r > r_0}
(\delta \hat A_{ab}) \Lie_{\Omega\phi} \tilde\gamma^{ab} d^3x. 
\eeqn
Since $(\delta \hat A_{ab})\Lie_{\Omega\phi} \tilde\gamma^{ab}$ falls
off as $O(r^{-4})$ and the integral is done only in the
distant zone, the magnitude of the integral would be very small. 
Thus, even with the modified formulation, the
first law would be satisfied approximately.
Furthermore, the virial relation is satisfied in this formulation. 

The condition for $\pa_t \hat A_{ab}$ may be changed to
\beq
v_{ab}=\left\{
\begin{array}{ll}
-\Lie_{\Omega\phi} \hat A_{ab}& {\rm for}~r \leq r_0, \\
0 & {\rm for}~r \geq r_0. 
\end{array}
\right.
\eeq
Then, 
\beqn
\delta \Madm - \Omega \delta J
&=&{1 \over 16\pi} \int_{r > r_0}
\biggl[(\delta \hat A_{ab}) \Lie_{\Omega\phi} \tilde\gamma^{ab}
-(\Lie_{\Omega\phi} \hat A_{ab}) \delta \tilde\gamma^{ab}\biggr]d^3x. 
\eeqn
Even in this case, the magnitude of the violation of the
first law would be small, and the virial relation holds. 
The merit in this approach is that the right-hand side
of the elliptic equation for $\thh_{ab}$ falls off as $O(r^{-4})$.
As a result, it is numerically easier to integrate the equation. 

We also note that instead of using the step function, we
may write
\beqn
&& u^{ab}=-\Lie_{f(r)\Omega\phi} \tilde \gamma^{ab},\\
&& v_{ab}=-\Lie_{f(r)\Omega\phi} \hat A_{ab},
\eeqn
where $f(r)$ is a smooth function that satisfies the condition 
\beq
f(r)=\left\{
\begin{array}{ll}
1 & {\rm for}\ r \ll r_0,\\
0 & {\rm for}\ r \gg r_0. 
\end{array}
\right.
\eeq
This choice is equivalent to taking a Killing vector of the form 
\beq
k^{\mu}=
\biggl({\pa \over \pa t}\biggr)^{\mu}
+f(r)\Omega \biggl({\pa \over \pa \varphi}\biggr)^{\mu}. 
\eeq
This Killing vector is helical in the near zone and purely timelike
for $r\rightarrow \infty$. 

Finally we comment on other possible formulations. 
In the formulation with $u^{ab}=0=v_{ab}$, the
virial relation is satisfied for a quasiequilibrium
binary. However, the first law along quasiequilibrium sequences
is not satisfied in general. The violation of the first law is written as
\beqn
\delta \Madm - \Omega \delta J
&=&{\Omega \over 16\pi} \int
\biggl[(\delta \hat A_{ab}) \Lie_{\phi}\tilde\gamma^{ab}
-(\Lie_{\phi} \hat A_{ab}) \delta \tilde\gamma^{ab}\biggr]d^3x \nonumber \\ 
&=&-{\Omega \over 16\pi} \delta \int
(\Lie_{\phi} \hat A_{ab}) \tilde\gamma^{ab} d^3x.
\eeqn

\section{Summary}

Two relations, the virial relation $\Madm=\MK$ and 
the first law $\dl\Madm=\Omega\dl J$, are regarded as guiding 
principles to develop a formalism for computing binary compact 
objects in quasiequilibrium circular orbits in general relativity.  
Deriving the explicit equations for $\Madm-\MK$ and 
$\dl\Madm-\Omega \dl J$ on the assumption that 
the spacetime is asymptotically flat, it is shown that a 
solution and a sequence of the solutions computed in 
some formulations satisfy these two conditions at least approximately. 
We propose a formulation in which the full Einstein equation is solved 
with the maximal slicing and in a transverse gauge
for the conformal three-metric. 
In the proposed formulation, the solution in the near zone is helically 
symmetric, but in the distant zone, it is asymptotically waveless. 

So far, quasiequilibria of binary neutron stars have been 
computed using the conformal flatness approximation for 
the three-metric \cite{BGM,UE}. In this formulation, only five 
components of the Einstein equation are satisfied,
and thus, the obtained numerical solutions for quasiequilibria
involve a systematic error. Specifically, in a real solution of
the quasiequilibrium circular orbit, 
the conformal nonflat part of the three-metric will be of order 
$(M/a)^2$, which can be $\sim 0.1$ near the neutron stars for 
close circular orbits of $a \alt 10M$ (e.g., \cite{SU01}).
This implies that to compute an accurate quasiequilibrium in circular orbits
of error within, say, 1\%, it will be necessary to take into account
the conformal nonflat part of the three-metric. 
In the new formulations described here, such term is computed, and
thus, more accurate solutions of quasiequilibria will be obtained. 
Currently, we are working in computation of binary neutron stars 
in quasiequilibrium circular orbits using these formulations.
In a subsequent paper \cite{Uryu}, we will present the numerical results.
Such a numerical solution will be also used as 
an appropriate initial condition 
for simulations of binary neutron star mergers \cite{SU}. 

In this paper, we restrict our attention to the system 
in which no black hole exist. In the presence of
black holes, we should carefully treat the
surface terms at event horizons. The surface terms would 
modify the equations for the virial relation and first
law \cite{FUS,cook}. The formulation for
computation of quasiequilibrium black hole binaries are
left for the future \cite{GGB02}. 

\begin{acknowledgments}
We are grateful to Greg Cook for helpful discussions and 
to Eric Gourgoulhon for comments on the manuscript.  
A part of this work was done when MS was visiting at Caltech in March 2003. 
MS is grateful to Kip Thorne and Lee Lindblom for their hospitality. 
This work is supported in part by Japanese Monbu-Kagakusho Grant
Nos. 15037204, 15740142, and 16029202 and by NSF Grant PHY0071044.
\end{acknowledgments}

\appendix

\section{Evolution equation for the scalar moment and virial relation}
\label{sec:virial}

In this section, we derive the virial relation by direct
integration of the Euler equation. Thus, 
the virial relation we consider here is associated with an evolution 
equation for the scalar moment as in the Newtonian case. In the end, 
we confirm that the virial relation derived is equivalent to $\MK=\Madm$. 

In the following, we often refer to $\Mchi$ as a ``Komar like mass'' 
that is defined by the asymptotic behavior of 
a function $\chi:=\alpha\psi$ at $r \rightarrow \infty$,
\beq
\chi \rightarrow 1 - {M_{\rm \chi} \over 2r} + O(r^{-2}). \label{bounda}
\eeq
For simplicity, in the following calculation, 
we adopt a gauge in which $K=0$, $F^k=0$, and 
$\tilde \gamma=\eta$, and we carry out the calculations in 
Cartesian coordinates. 
We often need to evaluate surface integrals at
$r \rightarrow \infty$. In the evaluation, we 
assume Eqs. (\ref{bound1})--(\ref{bound6}) as well as (\ref{bounda}) 
the asymptotic behaviors of geometric variables. 
As a consequence, all the volume integrals that appear below 
are well defined, and furthermore, 
the surface integrals derived during 
the calculation can be safely discarded. 

{}From the asymptotic behavior as $r \rightarrow \infty$, 
we can define $\Mchi$ and $\Madm$ using the surface integrals 
\beqn
\Mchi &=& {1 \over \2pi} \int_\infty 
\psi \pa_i \chi dS^i, \\
\Madm &=& -{1 \over \2pi} \int_\infty 
\chi \pa_i \psi dS^i. 
\eeqn
Using the Gauss's law, they can be rewritten in other forms 
\beqn
\Mchi &=& 
{1 \over 2\pi} \int (\psi \tilde \Delta \chi
+ \tilde \gamma^{ij} \pa_i \chi \pa_j \psi) 
d^3x, \label{eqmchi} \\
\Madm 
&=& -{1 \over \2pi} \int (\chi \tilde \Delta \psi
+\tilde \gamma^{ij} \pa_i \psi \pa_j \chi) d^3x. 
\label{eqmadm}
\eeqn
The difference between $\Madm$ and $\Mchi$ is written in the form 
\beqn
\Mchi - \Madm =&& \int \biggl[ 2 \chi \psi^5 S_k^{~k} 
+ {3 \over 8\pi} \chi \psi^5 \tilde A_{i}^{~j} \tilde A_{j}^{~i} 
+{1 \over 8\pi} \chi \psi \tilde R 
+{1 \over \pi} \tilde \gamma^{ij}\pa_i\psi \pa_j \chi
\biggr]d^3x. \label{eqdiff}
\eeqn
Here, using Eq. (\ref{heq}), we can derive an identity, 
\beqn
&&\int \alpha \psi^6 \tilde A^i_{~j} \tilde A^j_{~i} d^3x 
={1 \over 2}\int \psi^6 \tilde A^i_{~j}
(\pa_i\beta^j +\tilde \gamma_{ik}\tilde \gamma^{jn}\pa_n \beta^k
+\tilde\gamma^{jn}\beta^k\pa_k\tilde \gamma_{ni}-\tilde \gamma^{jk}
u_{ik})d^3x \nonumber \\ 
&&=\int \Big[ \psi^6 \tilde A^i_{~j} \pa_i\beta^j
+{1 \over 2}\psi^6 \tilde A^{jn} \beta^k \pa_k \tilde \gamma_{jn}
-{1 \over 2}\psi^6 \tilde A^{ij}
u_{ij} \Big] d^3x \nonumber \\
&&=\int \Big[ - \Big\{ \pa_i(\psi^6 \tilde A^i_{~j}) 
-\psi^6 \tilde A^{l}_{~k} \tilde \Gamma^k_{jl} \Big\} \beta^j
+{1 \over 2}\psi^6 \tilde A_{ij}
u^{ij} \Big]d^3x \nonumber \\
&&=-\int \Big[8\pi j_i \psi^6 \beta^i - {1 \over 2} \hat A_{ij}
u^{ij} \Big]d^3x, \label{id1} 
\eeqn
where $\tilde \Gamma^k_{ij}$ denotes the Christoffel symbol
with respect to $\tilde \gamma_{ij}$, and $\hat A_{ij}=\psi^6\tilde A_{ij}$. 
Thus, we obtain
\beqn
\Mchi - \Madm =&& \int \biggl[ \psi^6 j_l (2v^l-\beta^l)
+6\alpha \psi^6 P
+ {3 \over 16\pi} \hat A_{ij} u^{ij}
+{1 \over 8\pi} \chi \psi \tilde R 
+{1 \over \pi} \tilde \gamma^{ij}(\pa_i\psi) \pa_j \chi
\biggr]d^3x, \label{eqdiff2}
\eeqn
where we use the relation 
\beq
\alpha S_k^{~k}={j_k u_l \gamma^{kl} \over u^t} +3 \alpha P
=j_k (v^k +\beta^k) +3 \alpha P. 
\eeq
(Note that $v^k$ here is defined by $v^k=u^k/u^t$.) 
In stationary spacetimes, the relation $\Madm=\Mchi$ \cite{beig,ashtekar}
and $u_{ij}=\pa_t \tilde \gamma_{ij}=0$ should hold. Thus, we get the 
virial relation as 
\beq
\int \biggl[ \psi^6 j_l (2v^l-\beta^l) +6\alpha \psi^6 P
+{1 \over 8\pi} \chi \psi \tilde R 
+{1 \over \pi} \tilde \gamma^{ij}(\pa_i\psi)\pa_j \chi
\biggr]d^3x=0. 
\eeq
In quasiequilibrium binaries, $u_{ij}\not=0$ in general. 
Thus, the virial relation, $\Madm=\Mchi$, is written as 
\beq
\int \biggl[ \psi^6 j_l (2v^l-\beta^l) +6\alpha \psi^6 P
+ {3 \over 16\pi} \hat A_{ij} u^{ij}
+{1 \over 8\pi} \chi \psi \tilde R 
+{1 \over \pi} \tilde \gamma^{ij}(\pa_i\psi)\pa_j \chi
\biggr]d^3x=0. 
\label{eq:virial}
\eeq

As in the Newtonian case, we can derive the general relativistic
virial relation from the evolution equation for the scalar moment.
First, we write the general relativistic Euler equation 
$\gamma^{\nu}_{~k} \nabla_{\mu} T^{\mu}_{~\nu}=0$ in the form 
\beqn
\pa_t (j_k \psi^6)+\pa_j (j_k \psi^6 v^j) + \pa_k (\alpha \psi^6 P) 
+ \rhoH \psi^5 \pa_k \chi -(\rhoH + 2 S_l^{~l})\chi \psi^4 \pa_k \psi 
- \psi^6 j_l \pa_k \beta^l
+{1 \over 2} \chi\psi S_{ij}\pa_k \tilde \gamma^{ij}=0. \label{eqEuler}
\eeqn
Equation (\ref{eqEuler}) is a fully general relativistic 
expression, and no simplification is done.
Taking an inner product with $x^k$, we have
\beqn
&& \int x^k \biggl[
\pa_t (j_k \psi^6)+\pa_j (j_k \psi^6 v^j) + \pa_k (\alpha \psi^6 P) 
\nonumber \\
&&\hskip 1.5cm
+ \rhoH \psi^5 \pa_k \chi -(\rhoH + 2S_l^{~l})\chi \psi^4 \pa_k \psi
- \psi^6 j_l \pa_k \beta^l +{1 \over 2}\chi \psi S_{ij}\pa_k
\tilde \gamma^{ij}\biggr]d^3x =0. 
\label{eq:momeq}
\eeqn
In the following, we carry out the integral for each term separately.

(1) First term: 
\beqn
I_1 := 
\int x^k \pa_t (j_k \psi^6) d^3x ={d \over dt} \int x^k j_k \psi^6 d^3x. 
\eeqn
In the Newtonian limit, $j_k \psi^6 \rightarrow \rho v^k =\rho dx^k/dt$, and
thus, this term leads to half of the second time-derivative of the
scalar moment, i.e., $\ddot I_{kk}/2$ .

(2) Second and third terms: By integration by parts, 
we immediately find 
\beqn
I_2&&:=\int x^k \pa_j (j_k v^j \psi^6)d^3x=-\int j_k v^k \psi^6 d^3x,\\
I_3&&:=\int x^k \pa_k (\alpha \psi^6P)d^3x =-3 \int \alpha \psi^6 P d^3x.
\eeqn
In the Newtonian limit, $-I_2$ and $-I_3$ are the terms associated with 
kinetic energy and internal energy. 

(3) Fourth and fifth terms: 
Using Eqs. (\ref{psieq}) and (\ref{alpeq}), 
we can rewrite the combination of them as 
\beqn
\rhoH \psi^5\pa_k \chi-(\rhoH + 2S_l^{~l})\chi \psi^4 \pa_k \psi
={\tilde R \over 16\pi}\pa_k (\chi\psi)
-{1 \over 2\pi} \biggl[ (\Delta \psi) \pa_k \chi + 
(\Delta \chi) \pa_k \psi \biggr] 
-{\psi^{12}\tilde A_i^{~j} \tilde A_j^{~i} \over 16\pi}\pa_k \biggl(
{\alpha \over \psi^6}\biggr). 
\eeqn
Taking into account an identity, 
\beqn
\int \Big[(x^k \pa_k \psi)\Delta \chi +(x^k \pa_k \chi)\Delta \psi\Big]d^3x 
=\int \Big[\tilde \gamma^{ij} (\pa_i \chi)\pa_j \psi +x^k (\pa_i \chi)
(\pa_j \psi)\pa_k \tilde \gamma^{ij}\Big] d^3x, 
\eeqn
where we discard the vanishing surface integral terms, we find 
\beqn
I_4&&:=
\int x^k [\rhoH \psi^5\pa_k \chi-(\rhoH + 2S_l^{~l})\chi \psi^4 \pa_k \psi]
d^3x \nonumber \\
&&={1 \over 16\pi}
\int \biggl[\tilde R  x^k \pa_k(\chi\psi)
-8\Big\{ \tilde \gamma^{ij} (\pa_i\chi)\pa_j \psi
+x^k(\pa_k\tilde \gamma^{ij})(\pa_i \chi) \pa_j \psi \Big\}
- \psi^{12} 
\tilde A_i^{~j} \tilde A_j^{~i} x^k \pa_k \biggl(
{\alpha \over \psi^6}\biggr)\biggr]d^3x.
\eeqn

(4) Sixth term: 
\beqn
I_5 &&:=  -\int \psi^6 j_l x^k \pa_k \beta^l d^3x \nonumber \\
&&=-{1 \over 8\pi} \int x^k \pa_k \beta^l \biggl[
\pa_i (\psi^6 \tilde A_l^{~i}) +{1 \over 2}\psi^6 \tilde A_{ij}
\pa_l \tilde \gamma^{ij} \biggr] d^3x \nonumber \\ 
&&=-{1 \over 8\pi} \int \biggl[
-\psi^6 \tilde A_l^{~i}(x^k \pa_k\pa_i \beta^l + \pa_i \beta^l)
+{1 \over 2}x^k (\pa_k \beta^l)
\psi^6 \tilde A_{ij} \pa_l \tilde \gamma^{ij} \biggr] d^3x \nonumber \\ 
&&=-{1 \over 8\pi} \int \biggl[
(\pa_i\beta^l) x^k\pa_k(\psi^6 \tilde A^i_{~l})
+2(\pa_i\beta^l) \psi^6 \tilde A_l^{~i}
+{1 \over 2}x^k (\pa_k \beta^l)
\psi^6 \tilde A_{ij} \pa_l \tilde \gamma^{ij} \biggr] d^3x \nonumber \\
&&=-{1 \over 8\pi} \int \biggl[
(\pa_i\beta^l) x^k\pa_k(\psi^6 \tilde A^i_{~l})
-2\beta^l\pa_i(\psi^6 \tilde A_l^{~i})
+{1 \over 2}x^k (\pa_k \beta^l)
\psi^6 \tilde A_{ij} \pa_l \tilde \gamma^{ij} \biggr] d^3x. 
\eeqn
Here, let us evaluate the first term. Using Eq. (\ref{heq}), 
\beqn
I_5'&&:=-{1 \over 8\pi} \int 
(\pa_i\beta^l) x^k\pa_k(\psi^6 \tilde A^i_{~l})d^3x \nonumber \\
&&=-{1 \over 8\pi} \int 
(2\alpha\tilde A^l_{~i}-\tilde\gamma^{jl}\tilde\gamma_{ik}\pa_j\beta^k
-\tilde\gamma^{jl}\beta^k\pa_k\tilde\gamma_{ij}
+\tilde\gamma^{jl}u_{ij})
x^k\pa_k(\psi^6 \tilde A^i_{~l})d^3x  \nonumber \\
&&=-{1 \over 8\pi} \int \biggl[
(2\alpha\tilde A^l_{~i}
-\tilde\gamma^{jl}\beta^k\pa_k\tilde\gamma_{ij}
+\tilde\gamma^{jl}u_{ij})
x^k\pa_k(\psi^6 \tilde A^i_{~l})
+(\pa_j\beta^k)\psi^6\tilde A^i_{~l}x^m\pa_m (\tilde\gamma^{jl}
\tilde \gamma_{ik})\biggr] d^3x-I_5'.
\eeqn
Thus,
\beqn
I_5'&&=-{1 \over 16\pi} \int
\biggl[(2\alpha \tilde A^l_{~i}-
\tilde \gamma^{lj}\beta^k\pa_k \tilde \gamma_{ij}
+\tilde \gamma^{jl}u_{ij})\pa_m(\psi^6 \tilde A^i_{~l})
+(\pa_j\beta^k)(\tilde A^{ij}\pa_m \tilde \gamma_{ik}
+\tilde A_{kl}\pa_m \tilde \gamma^{jl})\psi^6\biggr]x^m d^3x \nonumber \\ 
&&=-{1 \over 16\pi} \int
\biggl[
{\alpha \over \psi^6}\pa_m (\psi^{12}\tilde A^l_{~i}\tilde A^i_{~l})
+(\tilde \gamma_{ij}\beta^k\pa_k \tilde \gamma^{jl}
 +\tilde \gamma^{jl} u_{ij})\pa_m(\psi^6 \tilde A^i_{~l})
+(\pa_j\beta^k)(\tilde A^{ij}\pa_m \tilde \gamma_{ik}
+\tilde A_{kl}\pa_m \tilde \gamma^{jl})\psi^6\biggr] x^m d^3x
\nonumber \\
&&=-{1 \over 16\pi} \int
\biggl[\biggl\{-\pa_m\biggl({\alpha \over \psi^6}\biggr)
\psi^{12} \tilde A^l_{~i}\tilde A^i_{~l}
+\beta^k\pa_k \tilde \gamma^{jl}\Big(\pa_m(\psi^6 \tilde A_{jl})
-\psi^6 \tilde A^i_{~l}\pa_m \tilde \gamma_{ij} \Big)
+\tilde \gamma^{jl}u_{ij}\pa_m(\psi^6 \tilde A^i_{~l})
\nonumber \\
&& \hskip 2cm
+(\pa_j\beta^k)(\tilde A^{ij}\pa_m \tilde \gamma_{ik}
+\tilde A_{kl}\pa_m \tilde \gamma^{jl})\psi^6\biggr\} x^m 
-3\alpha \psi^6\tilde A^l_{~i}\tilde A^i_{~l} \biggr] d^3x. 
\eeqn
As a result, 
\beqn
I_5 &&=
-{1 \over 16\pi}\int 
\biggl[-3\alpha \psi^6\tilde A^l_{~i}\tilde A^i_{~l}
-\psi^{12}\tilde A^l_{~i}\tilde A^i_{~l}
x^m \pa_m \biggl({\alpha \over \psi^6}\biggr)
+\beta^k (\pa_k \tilde \gamma^{jl})
\Big(\pa_m(\psi^6 \tilde A_{jl})
-\psi^6 \tilde A^i_{~l}\pa_m\tilde \gamma_{ij}\Big)x^m \nonumber \\
&&\hskip 2.5cm
+\tilde \gamma^{jl} u_{ij} x^m \pa_m(\psi^6 \tilde A^i_{~l})
+(\pa_j\beta^k)(\tilde A^{ij}\pa_m \tilde \gamma_{ik}
+\tilde A_{kl}\pa_m \tilde \gamma^{jl})\psi^6 x^m 
\nonumber \\
&& \hskip 2.5cm
+x^k (\pa_k \beta^l)\psi^6 \tilde A_{ij}\pa_l \tilde \gamma^{ij}
-2\beta^l(16\pi j_l \psi^6 -\psi^6 \tilde A_{ij}\pa_l\tilde \gamma^{ij})
\biggr]d^3x \nonumber  \\
&&= -{1 \over 16\pi}\int 
\biggl[-{3 \over 2}\hat A_{kl} u^{kl}
-\psi^{12}\tilde A^l_{~i}\tilde A^i_{~l}
x^m \pa_m \biggl({\alpha \over \psi^6}\biggr)
+\beta^k (\pa_k \tilde \gamma^{jl})
\Big(\pa_m(\psi^6 \tilde A_{jl})
-\psi^6 \tilde A^i_{~l}\pa_m\tilde \gamma_{ij}\Big)x^m 
\nonumber \\&&\hskip 2.5cm
+\tilde \gamma^{jl} u_{ij} x^m \pa_m(\psi^6 \tilde A^i_{~l})
+(\pa_j\beta^k)(\tilde A^{ij}\pa_m \tilde \gamma_{ik}
+\tilde A_{kl}\pa_m \tilde \gamma^{jl})\psi^6 x^m 
\nonumber \\
&&\hskip 2.5cm
+x^k (\pa_k \beta^l)\psi^6 \tilde A_{ij}\pa_l \tilde \gamma^{ij}
-\beta^l(8\pi j_l \psi^6 -2\psi^6 \tilde A_{ij}\pa_l\tilde \gamma^{ij})
\biggr]d^3x. 
\eeqn

(5) Seventh term: Using Eq. (\ref{aijeq}), 
we rewrite it as  
\beqn
{\chi\psi \over 2} S_{ij} x^k \pa_k \tilde \gamma^{ij} 
&&={1 \over 16\pi}x^k \pa_k \tilde \gamma^{ij}
\biggl[\chi\psi
\Big(\tilde R_{ij}+R^{\psi}_{ij}-{1 \over \alpha}D_i D_j\alpha \Big)
\nonumber \\
+&&\psi^6 \Big( -2\alpha \tilde A_{ik}\tilde A_j^{~k}
+\tilde D_i\tilde \beta^k \tilde A_{kj}
+\tilde D_j\tilde \beta^k \tilde A_{ki}
-{2 \over 3}\tilde D_k\tilde \beta^k \tilde A_{ij}
+\beta^k\tilde D_k \tilde A_{ij}-\pa_t \tilde A_{ij}
\Big)\biggr], 
\eeqn
where we use $\tilde \gamma_{ij}\pa_k\tilde \gamma^{ij}=
\pa_k \ln \tilde \gamma=0$.

By straightforward calculations, we obtain
\beqn
I_6&&:= {1 \over 16\pi}
\int \chi\psi x^k (\pa_k \tilde \gamma^{ij}) \tilde R_{ij} d^3x
=-{1 \over 16\pi}
\int \Big[\pa_l(\chi\psi)x^k (\pa_k \tilde \gamma^{ij})\tilde \Gamma^l_{ij}
+x^k\pa_k (\chi\psi)\tilde R + \tilde R \chi\psi \Big]d^3x,\\
I_7&&:= {1 \over 16\pi}
\int \Big[\chi\psi \tilde R^{\psi}_{ij}-\psi^2 D_i D_j \alpha \Big]x^k\pa_k
\tilde \gamma^{ij}d^3x
={1 \over 16\pi}\int \big[ -\tilde D_i \tilde D_j (\chi\psi)
+8(\pa_i \psi) \pa_j \chi \Big]x^k\pa_k \tilde \gamma^{ij} d^3x \nonumber \\ 
&& 
={1 \over 16\pi}\int \Big[ \tilde \Gamma^l_{ij} \pa_l (\chi\psi)
+8(\pa_i \psi) \pa_j \chi \Big]x^k\pa_k \tilde \gamma^{ij} d^3x .
\eeqn
Here, to derive the first equation, we use the spatial gauge condition
$F^k=0$ and relations in the present gauge as
\beqn
\tilde R = -{1 \over 2}(\pa_l \tilde \gamma^{ij}) \tilde \Gamma^l_{ij}
=\tilde \gamma^{ij} \tilde \Gamma^l_{ik} \tilde \Gamma^k_{jl}. 
\eeqn
The spatial gauge condition is also used in calculation for $I_7$. 

To evaluate the remaining terms, we
first rewrite the following equation using the definition of
$\tilde A_{ij}$ as
\beqn
&&-2\alpha \tilde A_{ik} \tilde A^k_{~j}
+\tilde D_i \beta^k \tilde A_{kj}
+\tilde D_j \beta^k \tilde A_{ki}
-{2 \over 3} \tilde D_k \beta^k \tilde A_{ij}
+\beta^k \tilde D_k \tilde A_{ij} \nonumber \\
&&=\beta^k\pa_k \tilde A_{ij}
-\tilde A_j^{~k}\tilde \gamma_{il}\pa_k \beta^l
+\tilde A_{ik} \pa_j \beta^k-\tilde A^l_{~j}
\beta^k\pa_k \tilde \gamma_{il}+\tilde A^k_{~j} u_{ik}. 
\eeqn
Then, after a straightforward calculation, we get
\beqn
I_8&&={1 \over 16\pi}\int\biggl[
-2\alpha \tilde A_{ik} \tilde A^k_{~j}
+\tilde D_i \beta^k \tilde A_{kj}
+\tilde D_j \beta^k \tilde A_{ki}
-{2 \over 3} \tilde D_k \beta^k \tilde A_{ij}
+\beta^k \tilde D_k \tilde A_{ij}-\pa_t \tilde A_{ij}
\biggr]\psi^6 x^l\pa_l\tilde \gamma^{ij} d^3 x\nonumber \\
&&={1 \over 16\pi}\int \biggl[
-(\pa_t \hat A_{ij})x^l\pa_l \tilde \gamma^{ij}
+\psi^6 \tilde A_{ij}(\pa_k\tilde\gamma^{ij}) x^l\pa_l\beta^k
+2\beta^k (\pa_k \tilde\gamma^{ij}) \tilde A_{ij}\psi^6
+\beta^k x^l (\pa_k\tilde \gamma^{ij}) \pa_l(\psi^6\tilde A_{ij})\nonumber \\
&&~~~~~~~~~~~~+\psi^6 x^l \tilde A^{ik}(\pa_k \beta^j)\pa_l \tilde \gamma_{ij}
+x^l\psi^6 \tilde A_{ik}(\pa_j\beta^k) \pa_l \tilde \gamma^{ij}
-\beta^k x^n \psi^6\tilde A^l_{~j}(\pa_k \tilde\gamma_{il})
\pa_n \tilde \gamma^{ij}+x^l \psi^6 \tilde A^k_{~j}
u_{ik} \pa_l \tilde \gamma^{ij}
\biggr]d^3x, 
\eeqn
where we use an identity $\pa_t \ln \psi^6 =D_k \beta^k
=\psi^{-6}\pa_k (\psi^6 \beta^k)$ that follows from 
the maximal slicing condition $K=0$.
Eventually, we find that $I_5+I_8$ has the following simple form
\beqn
I_5+I_8 && =
{1 \over 16\pi}\int \biggl[
-{3 \over 2}\hat A_{kl} u^{kl}
+\psi^{12} \tilde A_{i}^{~j}\tilde A_{j}^{~i}
x^n\pa_n\biggl({\alpha \over \psi^6}\biggr)
+8\pi j_k\psi^6 \beta^k
- v_{ij} x^n \pa_n \tilde \gamma^{ij} -\hat A_{ij} x^n \pa_n u^{ij}
\biggr]d^3x.
\eeqn

By summation of $I_1 \sim I_8$, we obtain the following simple relation: 
\beqn
0 =\sum_{i=1}^8 I_i 
&& = {d \over dt}\int x^k j_k \psi^6 d^3x 
-\int \biggl[ {1 \over 16\pi} \chi\psi \tilde R 
+j_k v^k \psi^6 + 3 \alpha \psi^6 P 
+{1 \over 2\pi}\tilde \gamma^{ij}\pa_i \chi \pa_j \psi  
-{1 \over 2} \psi^6 j_k \beta^k \nonumber \\
&&\hskip 4cm +{3\over 32\pi}\hat A_{ij} u^{ij}
+{1 \over 16\pi}
( v_{ij} x^n \pa_n \tilde \gamma^{ij} +\hat A_{ij} x^n \pa_n u^{ij})
\biggr] d^3x \nonumber \\
&&= {\Madm -\Mchi \over 2}+{d \over dt}\int x^k j_k \psi^6 d^3x
-{1 \over 16\pi} \int
( v_{ij} x^n \pa_n \tilde \gamma^{ij} +\hat A_{ij} x^n \pa_n u^{ij}) d^3x
\label{simI}
\eeqn
Here, since $u^{ij}=\pa_t \tilde \gamma^{ij}$
and $v_{ij}=\pa_t \hat A_{ij}$, 
the second integral term in the last line of Eq. (\ref{simI})
is rewritten as
\beq
-{1 \over 16\pi} {d \over dt} \int
( \psi^6 \tilde A_{ij} x^n \pa_n \tilde \gamma^{ij} ) d^3x.
\eeq
Using the momentum constraint, we further rewrite this term as
\beqn
&&-{1 \over 16\pi}
{d \over dt} \int
 ( \psi^6 \tilde A_{ij} x^k \pa_k \tilde \gamma^{ij} ) d^3x \nonumber \\
=&&-{1 \over 8\pi}{d \over dt} \biggl[
\int x^k \biggl\{\pa_i (\tilde A^i_{~k}\psi^6)
+{1 \over 2}\psi^6 \tilde A_{ij}\pa_k \tilde \gamma^{ij}\biggr\}d^3x
-\int_\infty
dS_i \tilde A^i_{~k} x^k \psi^6 \biggr] \nonumber \\ 
=&&-{d \over dt} \biggl[
\int x^k j_k \psi^6 d^3x -{1 \over 2}Z_{kk}  \biggr]. 
\eeqn
Thus, a similar relation between $\Madm$ and $\Mchi$ 
\beq
\Madm = \Mchi - {d Z_{kk} \over dt}. \label{virial}
\eeq
is derived as is done for $\Madm$ and $\MK$ in Sec.~\ref{sec:Madm=MK}. 
In this way, one can associate a relation of two masses
$\Madm$ and $\Mchi$ ($\MK$) to a moment equation of the relativistic 
Euler equation Eq.~(\ref{eq:momeq}).   

In the Newtonian theory, we usually check the accuracy of numerical 
solutions by the virial relation. 
Since the relation is not trivially satisfied in numerical solutions, 
violation of this relation can be 
used to estimate the magnitude of the numerical error of equilibria. 
Motivated by this idea, a virial relation is also derived for 
axisymmetric equilibrium states in general relativity
\cite{BG}, and it is subsequently 
used to check accuracy of numerical solutions for 
rotating neutron stars \cite{NSGE}.
The virial relation has been also derived for
binary neutron stars in quasiequilibrium in conformally flat
spacetimes \cite{FUS} and applied for monitoring accuracy 
of numerical solutions in \cite{TG0203}. 
The virial relation, e.g., Eq.~(\ref{eq:virial}), derived here will be used 
when checking the accuracy 
of nonaxisymmetric numerical solutions.

\end{document}